\documentclass[12pt]{article}
\usepackage{authblk}
\usepackage{moreverb}
\usepackage[super]{cite}
\usepackage{graphicx}
\usepackage{url}
\usepackage{amsmath,bm}
\begin{document}

\title{Analysis of the U.S. Patient Referral Network}

\author[1]{\small Chuankai An*}
\author[2]{\small A. James O'Malley}
\author[1,3]{\small Daniel N. Rockmore}
\author[3]{\small Corey D. Stock}

\affil[1]{\footnotesize Department of Computer Science, Dartmouth College, Hanover, NH, 03755, USA}
\affil[2]{\footnotesize Department of Biomedical Data Science and the Dartmouth Institute of Health Policy and Clinical Practice, Geisel School of Medicine at Dartmouth, Lebanon, NH, 03756, USA}
\affil[3]{\footnotesize Department of Mathematics, Dartmouth College, Hanover, NH, 03755, USA}


\maketitle

\begin{abstract}
In this paper we analyze the \emph{US Patient Referral Network} (also called the \emph{Shared Patient Network}) and various subnetworks for the years 2009--2015. In these networks two physicians are linked if a patient encounters both of them within a specified time-interval, according to the data made available by the Centers for Medicare and Medicaid Services.  We find power law distributions on most state-level data as well as a core-periphery structure. On a national and state level, we discover a so-called small-world structure as well as a ``gravity law" of the type found in some large-scale economic networks. Some physicians play the role of hubs for interstate referral. Strong correlations between certain network statistics with healthcare system statistics at both the state and national levels are discovered. The patterns in the referral network evinced using several statistical analyses involving key metrics derived from the network illustrate the potential for using network analysis to provide new insights into the healthcare system and  opportunities or mechanisms for catalyzing improvements.
\end{abstract}


\section{Introduction and Background}\label{sec:intro}

A well-designed healthcare system is a key component of a working society, and the ability for information and resources to flow efficiently in such a system is crucial to its efficacy. Inefficient communication in the United States healthcare system oftentimes results in ineffective care coordination.~\cite{Zuchowski2015} Referrals are one of the most common and important forms of primary-specialty care communication. The existence of a  shared patient relationship between physicians likely means there exists professional, information-sharing relationships between the two.~\cite{barnett2011mappingthesis,uddin2013mapping} Physicians decide to refer patients to other physicians in other hospitals for a multitude of reasons ranging from the need for specialization to addressing problems of overcrowding. The top three reasons for a primary care physician to refer a patient to a specialist are (1) seeking advice on a diagnosis or treatment (52.1\%), (2) requesting surgical management (37.8\%), and (3) asking the specialist to directly manage the patient (25.1\%).~\cite{Forrest2002} A physician's decision to refer (or not refer) a patient is important in determining cost and quality of care.~\cite{barnett2012reasons} See also.~\cite{barnett2011mapping,barnett2012trends}

The referral of a patient by physician $A$ to physician $B$ is naturally represented as a directed edge from a network node labeled  $A$ to a node labeled $B$, forming a \emph{directed network} (possibly weighted by the number of such referrals).~\cite{barnett2011mappingthesis} (See https://blog.pokitdok.com/physician-co-occurrence-graphs-part-1-of-2/ for visualizations of physician ``ego networks": network diagrams illustrating all referral relationships that include a single particular physician.) In this paper we analyze the structure of an important physician network and in this context introduce a number of novel concepts from the network science and social networks fields; drawing together methods from both of these two growing but surprisingly distinct fields is an important and novel feature of this work that is hoped will catalyze their combined use across the statistical sciences as applied to healthcare related networks. 

In particular, in this paper we use data from the Centers for Medicare and Medicaid Services (see \cite{SharedPatientData,SharedPatientDataDesc}) that is freely available and therefore among (often closely held) healthcare data sets is uniquely amenable to enabling reproducible research. The key data element is the counts of patients encountered by one physician and then another physician within an interval of time (we use 30 days). The networks constructed from physicians' caring for the same patients have been variously described as a referral network, a collaboration network, an informal physician network, and a patient-sharing network. It reflects collaboration, coordination, communication channels, diffusion of information, and possibly diffusion of medical innovations.~\cite{uddin2013mapping,barnett2012reasons,barnett2011mapping,barnett2012trends,landon2012variation,Mand:2014} Although counts of shared patients do not necessarily represent formal referrals from one physician to the other, for ease of terminology we take {\it referral} to mean the event that a patient encounters the first physician followed by the second within 30 days.

One motivation for studying physician networks stems from reported associations of collaboration with effectiveness in delivering health services and its recognition as a catalyst to improved patient outcomes. \cite{Sawy:2010,Uddi:2012} For example, associations have been found between collaboration and hospital length of stay, lower re-admission rates, lower hospital cost, lower death rate, and higher satisfaction. \cite{Cowa:2006,Knau:1986,Poll:2014,Tsch:2009} A multi-level regression model~\cite{Uddi:2016} has previously been used to study the relationship between patient-care networks and healthcare outcomes. Our work differs from this in that we summarize networks using a broader range of metrics, including many from the field of network science and even measures such as gravity that may be unfamiliar to social network analysts, and use state-level healthcare metrics as opposed to patient-level metrics. Network analysis has the potential to differentiate physicians' local network structure and network positions, reflecting different structures of collaboration, and determining if these explain variation in important health outcomes. \cite{Chuk:2011} In so doing, network analysis can identify favorable network structure and provide a recipe for organizing healthcare in a more optimal way.

Prior studies have not provided guidance about the network structure of effective healthcare collaboration, i.e., they have not been able to articulate what types of structure may be  more conducive for the administration of effective healthcare. Nor have they prescribed how individual healthcare professionals should develop relationships over time for better outcomes. \cite{uddin2013mapping} Yet, the formation of physician networks from patient encounters as reported in health insurance claims data is becoming increasingly common. To date, the focus has been on the production of descriptive measures of the networks, including: density, degree, centrality (betweenness, eigenvector, Bonacich), strength (number of shared patients), clustering coefficient, assortativity, reciprocity, transitivity, cyclic triads, network distance and the positioning of physicians within the network (e.g., \cite{landon2012variation,Mand:2014,Lee:2011,Lomi:2014}). Much less focus has been paid to developing and testing generative models able to capture  the change in the network across time. Furthermore, the complexity of network data has often been overlooked and network statistics are often reported without statistical tests to determine if the feature in question is significantly different from what would be expected under a simple generative model of the network representing a meaningful null hypothesis. 

Another overlooked issue is the impact of artificially imposed boundary definitions on the network. We are uniquely positioned to investigate boundary effects given that our data covers the complete US referral network. This allows us to assess whether analyses of geographically-defined sub-networks (e.g., state networks) and derived structural assessments  are sensitive to the definition of  the boundary and thus may distort the relationship of the network to important health variables. 

A core area of research in health services research is the examination of the volume-outcome hypothesis under which physicians and hospitals who perform more procedures get better at them as reflected in their patient outcomes. A network-centric point of view suggests natural generalizations. If increased volume leads to better outcomes one might think that in a network context that the most well functioning healthcare organizations or regions will be those with the strongest physician network ties. However, the weak ties hypothesis \cite{Gran:1973,Iwas:2009} may imply a more complex story. Perhaps, rare referral pathways are the most crucial elements of the physician network. The analysis herein may be useful for exploring such hypotheses. 

We analyze the structure of patient referral networks at both national and state levels. We evaluate both macro (global) and micro (local configuration or actor specific) network features, describe the network in static and dynamic terms, and test against and for generative models such as the random network, the small-world, and power-law network, while also measuring the degree to which structural phenomena such as high core-periphery tendency are evident in the referral network. With longitudinal referral data between 2009 and 2015, we also are able to study changes in the connectivity of the healthcare system. Moreover, we incorporate healthcare metrics for the fifty states and apply regression models to estimate the relationship of state-level network features to them. The main contributions of our work are:
\begin{itemize}
  \vspace{-0.1in}
  \item Novel application of combined network science and social network methods to the referral network of all US physicians caring for patients on Medicare. The results show several revealing patterns in network structure.
  \vspace{-0.1in}
  \item Evaluation of whether national-level patterns and phenomena reproduce at the state-level and whether state-level, or other sub-network analyses may yield misleading results.
  \vspace{-0.1in}
  \item Estimation of multi-level regression models that extract the portion of the association between the network features and state level healthcare attributes that is independent of other predictors in a multiple regression framework.
\end{itemize}
It is important to note that the conclusions we draw are specific to the data source -- that is the population of patients using Medicare and the physicians who care for them. Nevertheless, even with that caveat, the novelty of our work is seen in the scale of analysis, the consideration of network metrics relevant to the question of whether particular generative stories for the network are plausible, and the investigation of whether both macro (whole network) and micro (actor specific) network metrics are associated with various healthcare and health litigation metrics. Previous research on patient referral networks considers different questions and with not nearly as complete or expansive a network as that considered herein. For example, Lee~\cite{lee2011social} focuses on a single hospital in Orange County, CA, showing a  tendency of physicians there to share patients with physicians at great distance (the ``weak ties hypothesis").  Barnett et al. study primary care physician referral patterns using logistic regression~\cite{barnett2012reasons} and find a relationship between patient sharing and at least one of the physicians reporting a professional relationships between them (via self-reported survey response), thereby providing some validation for using health insurance claims to identify patient-physician encounters to  develop subsequently physician networks. This patient-sharing basis for developing the network is the same as that used herein. Landon et al. \cite{landon2012variation} consider the patient referral network at the HRR level and show the variation in network features at this scale, while discussing the factors related to building the referral connection between physicians. In an important longitudinal investigation, Barnett~\cite{barnett2012trends} finds changes in trends around  referrals between physicians in 1999--2009 and that the structure of an HRR shared patient network has significant associations with costs and intensity of care. 
Landon et al. ~\cite{landon2012variation}   examines how physicians' professional networks differ across geographic regions. Referral networks can also be viewed as collaboration networks for physicians, making relevant the large body of work on collaboration networks (see e.g., \cite{Newman2001}) and networks of  business relationship such as  ``interlocking corporate boards".  \cite{PettigrewInBdNet}

The organization of the remainder of the paper is as follows. Section~\ref{sec:background} describes the data and outlines the methods and analyses used in this paper. Section~\ref{sec:result} presents our findings on the structural metrics of the referral networks. In Section~\ref{sec:healthcare} we relate these to healthcare statistics. Finally, Section~\ref{sec:conclusion} concludes by describing the potential of a combined approach encompassing generative and empirical models involving networks and their key features to provide insights into the US healthcare system that could suggest pathways toward improving its organization.
\vspace{-0.2in}

\section{Materials, Notation, and Methodology}
\label{sec:background}

\subsection{Materials}

\noindent {\bf Data.} We used the CMS patient referral data set \cite{SharedPatientData} to form a physician network of the US healthcare system. Datasets are available for the years 2009--2015, measuring the number of patients encountered by one physician and then the other physician within 30-, 60-, 90-, and 180-day interval per year, so the referrals are derived with a threshold (e.g. 30-day) over a given year. The referral dataset includes two parts, the referral records between two physicians and the attributes of the physicians. Sharing (referral) occurs when the same patient is recorded as having been treated by two different physicians in a given time period. The dates of treatment are timestamps. In this paper, we choose the 30-day interval referral dataset, because it judges the existence of direct referrals between two physicians with the most stringent criteria. For example, if a patient visits physician A two months after a visit to physician B the record will not be counted in the 30-day dataset, but will be counted in the  datasets with longer time window. A referral dataset with longer interval will include more referrals, so some patterns (e.g., ``small-world'' below) in the referral networks found for the 30-day interval dataset would only be strengthened for the more expansive dataset. Physicians are listed according to National Provider Identification (NPI) number. (There are $4,332,951$ physicians in the NPI dataset.) A full data description can be found online. \cite{SharedPatientDataDesc}  In addition, the National Bureau of Economic Research ``National Provider Identification number by state" data was used to attribute each physician in each year to a state based on their NPI. Some physicians are registered in several states. We label a physician according to the state in which the physician makes the most referrals.  Table~\ref{table:ref-data} shows the number of 30-day-interval referral records for the years 2009-2015. Notice that there are fewer referrals in 2015 due to the fact that data was only obtained for 7 months of the year (the end-date of the data is 10/1/2015 and so the last date for a first visit under which a full 60-days is available for a second visit is 7/31/2015). Accordingly, we expect a reduction in the average number of referrals between two physicians in 2015 compared to the earlier years. (This may mean that network features in 2015 might not be comparable to those in 2009-2014.)

\noindent {\bf Networks of interest.} To produce a network description of the data,  each physician is  a \emph{node} in the network. An \emph{edge} (\emph{link}) between two nodes (physicians) reflects the existence of the patient referral, and can be \emph{weighted} by the number of referrals or simply given a weight of $1$ to indicate the  connection.  The edge can be \emph{directed} (indicating time -- referral) or \emph{undirected} (encoding only sharing). 

Given this data we  form three kinds of networks (over a given time period): (1) The \emph{National Patient Referral Network} includes all physicians in the US who have either made or received referrals over the period; (2) The (50) \emph{State Patient Referral Networks} wherein for state $S$, the node set is all physicians who are either labeled as physicians in state $S$ or have either made referrals to or received referrals from physicians labeled with state $S$ over the period; The (50) \emph{Intrastate  Patient Referral Networks}  is a subnetwork of the  State Patient Referral Networks and requires that both physicians in a referral be labeled as in state $S$. The node set for state $S$ is all physicians with NPI numbers in state $S$ who have either made or received referrals over the given period. In network terminology the State Patient Referral Network would be called the subnetwork \emph{induced} by the Intrastate Patient Referral Network. The three kinds of networks are nested as 

$$\begin{array}{c}
\mbox{National Patient Referral Network}\cr
\cup\cr
  \mbox{(Induced) State  Patient Referral Network of State }  S \cr 
   \cup \cr
\mbox{Intrastate State  Patient Referral Network of State } S 
\end{array}$$
for each state $S$. Each of these networks can be studied as simple undirected or directed networks, weighted or unweighted  (wherein the weights are the number of referrals). These networks are also called \emph{Shared Patient Networks}.

\subsection{Macro-level and small scale network structures}

There are still a relatively small number of well-defined -- or at least named -- macro-level structures. Three of interest are the {\rm random},  \emph{small world} and \emph{core-periphery} networks. We first introduce these and then describe various small scale or local network structures of interest whose prominence in the network can be tested against assumed models of macro-level structure.

\begin{figure}[h]
  \vspace{-0.21in} 
  \centering
    \includegraphics[height=1.5in,width=2.5in]{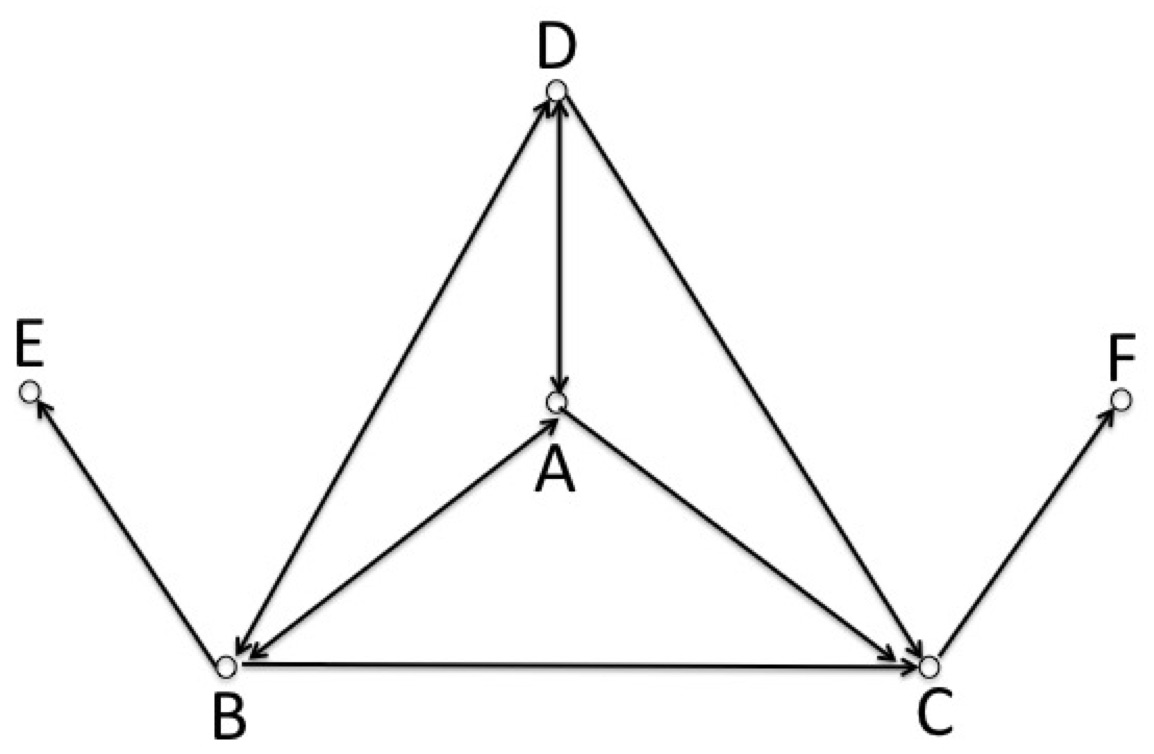}
  \vspace{-0.11in}
  \caption{An illustrative directed network. The nodes A, B, C, D, E, and F  represent different physicians. The arrow of an edge points from a referring physician to a referred physician (who accepts the patient referral).}
  \vspace{-0.22in}
  \label{fig:graph-notion} 
\end{figure}

\begin{itemize}

\item{Erd\'os-Renyi (ER) random network} -- is the traditional null model against which network structure is measured. The ER network on a fixed number $n$ of nodes is constructed by independently joining any two vertices with an (undirected) edge with fixed probability $p$. \cite{Erdos} It is easy to see that the expected degree for any vertex in such a network is $\mu=(n-1)p$, and that the degree distribution follows the binomial $B(n-1,p)$, which for large $n$ is well approximated by a Poisson distribution with mean $\mu$. Ascribing structure to a network derives from showing that in various important parameters it differs from the comparable ER network with probability $p = \mu/(n-1)$ where $\mu$ is the average degree of the actors in the network. 
 
 \medskip
 
\item{Small world network} --  is defined as a network with greater than expected local connectivity and average path length \cite{WattsStrogatzSW} smaller than expected in a comparable ER random network. More rigorously, a network is a small world if it has a higher (local) \emph{clustering coefficient} (defined as the average tendency for two ``friends of a common friend  to be friends themselves", cf. below)  and much smaller characteristic path length than expected under the Erd\'os-Renyi random graph model.  The work in \cite{WattsStrogatzSW} kicked off an era of discovery of small world structure in a wide range of naturally occurring networks. As we will see, some of the patient referral networks have small world characteristics, which suggests that a common organizational structure  for healthcare may be  found (or have evolved) irrespective of the size of the state (or possibly any other health unit). 

\medskip

\item{Core-Periphery structure} -- is a generative network model whose departure from the ER model is due to the network containing a ``core" subset of interconnected nodes, which are also connected to a less interconnected subset of ``peripheral" nodes. \cite{yang2014overlapping} For instance, in Figure~\ref{fig:graph-notion}, A, B, C, D are core nodes with connections to a collection of neighbors, while E and F are peripheral nodes. 
A core-periphery structure might occur in healthcare if the practice of medicine is primarily driven by a subgroup of inter-connected physicians that impart tremendous influence. By comparison, it might be that some states have a more uniform network in which there exists no such subgroup. The ``core-ness" of a node can be quantified via the assignation of a \emph{Core-Periphery (CP)} score  to each node. \cite{rombach2014core} The range of the CP score is $[0, 1]$, with $1.0$ indicating the node has the highest core quality. 
The extent to which a network has a generalized star structure can be captured by  the Gini coefficient (cf. \cite{GiniDef}) of the set of CP scores in the network. This is a standard measure of dispersion in a collection of numbers. 

\medskip 

\end{itemize}

There are various structural metrics fundamental to describing any network (see \cite{NewmanReview,Omal:2013} and the references therein) and so will be important for our analysis. The presence of a particular structural feature or phenomenon is ideally discovered by claiming that the observed network is highly unlikely to have arisen under a null model that exchanges randomness for the structural feature in question. In practice, investigators often claim that their network exhibits a certain trait by using the   Erd\'os-Renyi (ER) network as a null model. Such a comparison risks confounding the feature in question with any other feature that is not controlled. The distributional comparisons made in this paper are limited to single feature departures from the ER network. With this in mind, we describe various measures of small-scale network structure used herein and describe statistical tests of the extent of their prominence in the network beyond that expected by chance.

\begin{itemize}
\item{Degree Statistics} -- in an undirected network, the \emph{degree} of a node is the number of edges incident to the node, which is the same as the number of \emph{neighbors} of the node, or in the referral networks, the number of distinct physicians that a given physician has referred to (shared patients with) and/or received referrals from. If the edges are weighted, degree is the sum of the incident weights. In a directed network, there is an \emph{indegree} and an \emph{outdegree}, counting the number of edges coming \emph{into} the node (number of physicians that refer \emph{to} a given physician) and the number of edges issuing \emph{from} the node (number of physicians that a given physician refers \emph{to}). In Figure~\ref{fig:graph-notion}, for node A, indegree is 2 while outdgree is 3. The indegree of Node F is 1 and outdgree is 0. If the edges are weighted, the in- or outdegree is computed by summing over the weights of incoming or outgoing edges, respectively. The \emph{degree distribution} is the frequency distribution of the degrees (analogously for the in- or outdegree distribution). 

Various families of degree distributions appear in the network literature. As mentioned,  the undirected \emph{Erdos-Renyi random network} produces  an expected degree-distribution that is a binomial distribution with probability parameter equal to the proportion of non-null ties. Asymptotically, as the number of referrals increases, the degree distribution will converge to Poisson. However, over the past decade or so, much attention has been paid to kinds of ``heavy-tailed" distributions, especially those that follow a \emph{power law} 
\begin{equation}
\label{equal:1}
 y = Cx^{-\alpha}. 
\end{equation}
that are often found in data. Power laws can arise for a number of reasons (see \cite{MitzPL,NewmanPL}) and their discovery in data is but a starting point for a deeper investigation into an appropriate generative model.  Measurement of a power law can be subtle. We use the estimation method in \cite{clauset2009power} and perform calculations in $R$. \cite{poweRlaw} 

\medskip

\item{Cluster coefficient} -- a \emph{cluster coefficient} measures the extent to which nodes cluster together in a network. It is a measure taken on undirected networks of the frequency with which a ``$3$-chain" -- defined as a triple of connected nodes (A is connected to B is connected to C) is completed to a ``triangle" (A is connected to B is connected to C is connected to A). The triple (A, B, C) in Figure~\ref{fig:graph-notion} constructs a ``triangle'' when the graph is treated as undirected since any two of them are directly connected, but without an edge between A and F, the triple (A, C, F) is only a ``connected'' triple rather than a ``triangle''.
\emph{Global clustering} $C_g$ measures the fraction of completed triangles over the entire network while \emph{local clustering} $C_l$ measures the average number of triads centered at a given node that are completed to triangles.
In social network terminology,  $C_l$ measures the average tendency for ``friends of individual $i$ to also be friends of each other". If the network is homogeneous such that $C_{l}$ is invariant to $l$ then $C_{l}=C_{g}$ for $l=1,\ldots,n$. Under the ER network, the expected value of $C_{g}$ equals the probability of a single edge ("density"), denoted $p=\sum_{i} k_{i}/(n(n-1))=\mu/(n-1)$, where $\mu$ denotes average degree across the network, with standard error SE$ = p(1-p)/(n(n-1))$. High deviation from this null in an actual network is one of the conditions of the small world structure (see above). This result allows for inference about $C_{g}$ although in a practical application a null based on a richer network than the ER will often enable a more specific conclusion.

\medskip

\item{Assortativity, Degree Distribution Correlation, Reciprocity} -- Various  kinds of measures of connectivity can be supplemented by measures that get at \emph{assortativity}, a general term for quantifying the degree to which ``likes link to likes" (also called \emph{homophily} in the  social network literature) where ``like" can refer to any kind of metadata. An intrinsic kind of assortativity in any network is \emph{degree assortativity}, often referred to as simply ``assortativity". It measures the predilection of high degree nodes to attach to other high degree nodes and low degree to low degree.  In directed networks there are thus four different kinds of degree assortativity:  (in-, in-), (in-,out-), (out-,in-), and (out-,out-) depending on which kind of degree is taken into account.  Let $e_{AB}$ represent the weighted edge from node $A$ to node $B$ in Figure~\ref{fig:graph-notion}, $A_{in}$ be the in-degree of node $A$ and likewise define $B_{in}$. In this example, $A_{in}=2$ and $B_{in}=2$ and there are two possible indegree values of the two edge nodes. Therefore, the (in-, in-)-assortativity can be described in terms of the Pearson correlation coefficient between those two values for all edges. Since an edge from  $A$ to $B$ does not necessarily mean there is another edge from  $B$ to $A$, $corr(A_{in}, B_{out})$ is not equal to $corr(A_{out}, B_{in})$. A large assortativity means physicians in the network tend to build connections to others who have similar degrees.

 \emph{Self-Degree Correlation} measures the correlation of in- and outdegree on the node level (measuring the relatedness between the number of referrals made with the number of referrals received).  For those nodes in Figure~\ref{fig:graph-notion}, the in-degree (e.g. $A_{in}$ =2) might be in accordance with the out-degree (e.g $A_{out}$=3). While assortativity describes the relationship of two nodes on the same edge, self- (in- and out-) degree correlation is evaluated as the nodes' in- and outdegree.

  Finally,  \emph{reciprocity} measures the pairwise relationship between two individual physicians, computing the correlation of \#referrals from $A$ to $B$ and $B$ to $A$, where physicians $A$ and $B$ are connected with bidirectional edges in the referral network. It reflects the extent of quid pro quo in patient referrals between two physicians. 

  Because the assortativity, self-degree correlation, and reciprocity statistics are based on correlation coefficients the null value for many statistical tests involving them will be 0. Under the ER network the standard-error is inherited from that for a correlation coefficient. However, under more complex null models for the network, the calculation is more complicated and a permutation test or some other numerical method might be needed so that the aspects of the network that hold under the null model are fixed.
\vspace{-0.09in}
\medskip
\item{Motifs} -- the physician-physician relationship is the core atomic structure of the referral network. Nevertheless, it makes sense -- and is often useful -- to attempt to identify other regularly repeating evolved substructures.~\cite{Omal-2008} Such subnetworks are called \emph{motifs}. Two node or dyadic motifs include null-dyads, directional dyads (e.g., Node B and E in Figure~\ref{fig:graph-notion}) and bidirectional or mutual dyads (e.g., Node A and D in Figure~\ref{fig:graph-notion}). A familiar example in an undirected network is the ``triangle" representing the common phenomenon that ``a friend of your friend is your friend". If one accounts for ``enemies" in a social network (say by allowing friendships to come with a ``sign"), then the motif of two friends having a common enemy (a particular signed triangle) is also common. In the case of directed referral networks, we are interested in exploring the landscape of small (three-node) motifs, or ``triads". In a directed network there are $16$ non-isomorphic kinds of triads  (cf. Figure~\ref{fig:Result-16triads}). Some researchers name them by the number of mutual, asymmetric and null dyads. We describe the distribution of the 16 triads across the physician network and use factor analysis to group the triad types into categories that can be represented more parsimoniously in regression models.
\vspace{-0.12in}
\end{itemize}
\subsection{Experimental settings}
Though the referral records are clean and complete in 2009-2014, in the NPI dataset some physicians belong to multiple states. To extract the state level subnetworks, we assign each of those physicians the state where they have the most connections.

Our computations are implemented in Python and R.  For some network features such as Core-Periphery (CP) score and diameter, we apply parallel computing with sampling to speed up the computation and in some cases make the computation feasible. 
\vspace{-0.05in}

\section{Results: Network Statistics}
\label{sec:result}
In this section~\ref{sec:result} we explore the network metrics described above.  Table
~\ref{table:ref-data} records the sizes of the networks.

\subsection{Network models}
\noindent {\bf{Small Worlds.}} ``Small world" networks   \cite{WattsStrogatzSW} are networks with greater than expected  (local) clustering coefficient and average path length (the average of the shortest sequence of edges -- ``shortest path" -- between each pair of path-connected nodes in the network) much smaller than expected under the comparable Erd\'os-Renyi random graph model \cite{Erdos} (see above). Computing average path length exactly is prohibitive, but as a proxy we use the longest shortest path in the network (the ``diameter") since no shortest path can be longer than the maximum.   The ER network has expected local clustering given by the network density, $p=\mu/(n-1)$ for $\mu$ average degree and $n$ nodes. As we will see, the patient referral networks have small world characteristics. 

\begin{figure}[h]
  \vspace{-0.24in} 
  \centering
    \includegraphics[height=1.4in,width=2.4in]{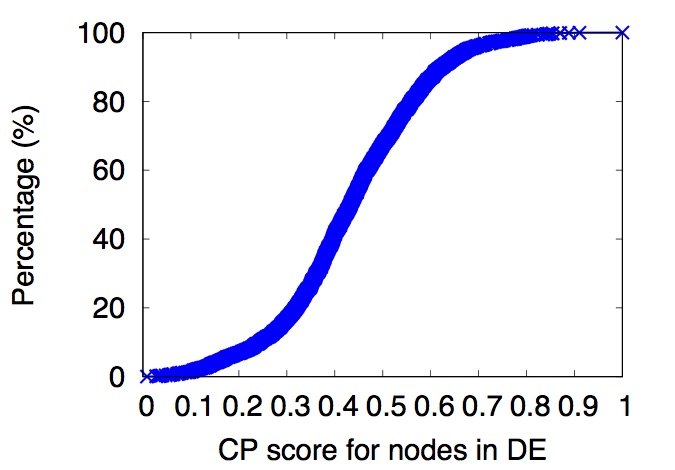}
	\hspace{-0.1in}
    \includegraphics[height=1.4in,width=2.4in]{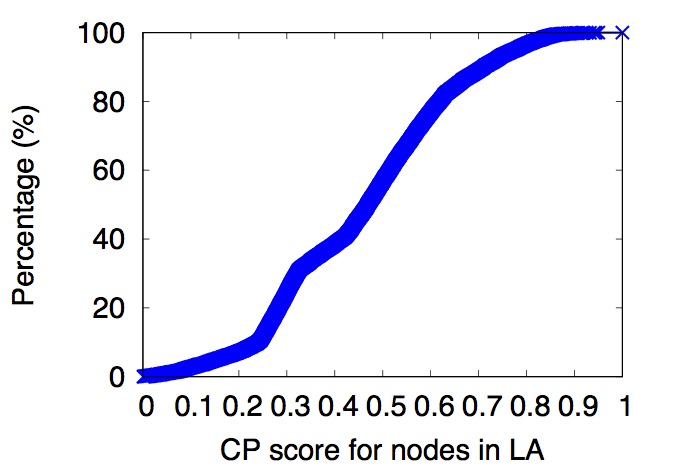}
	\hspace{-0.1in}
    \includegraphics[height=1.4in,width=2.4in]{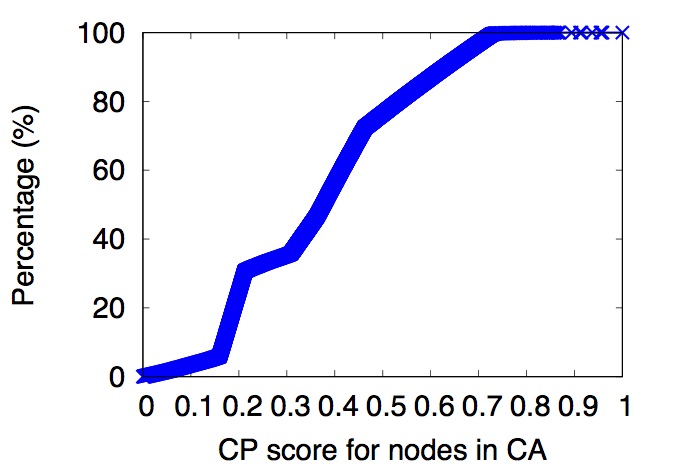}
	\hspace{-0.1in}
	\includegraphics[height=1.4in,width=2.4in]{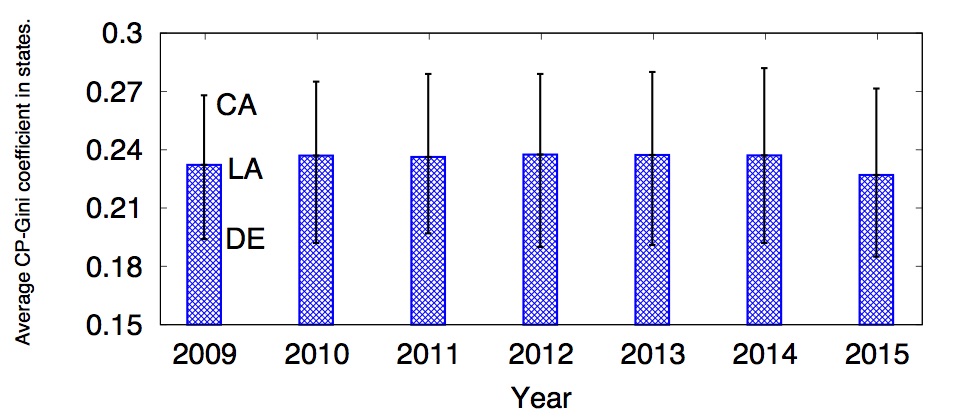}
  \vspace{-0.03in}
  \caption{\small{Counterclockwise from upper right: CP score distribution of LA, DE, and CA (minimu, median, maximum), in 2009 and the distribution of Gini coefficients of CP score  among the 50 states over 2009-2015.}}
  \vspace{-0.16in}
  \label{fig:CPS-gini-distribution} 
\end{figure}

\noindent {\bf{Core-Periphery Structure.}} Figure~\ref{fig:CPS-gini-distribution} gives an example of the CP (core-periphery) score distribution for the intra-state networks for states of DE, LA and CA  in 2009. These states werepicked because they have the minimum, median, and maximum of the Gini coefficients for the CP scores in 2009. Recall that a large Gini coefficient of the CP scores implies the network has a strong Core-Periphery structure: there are a small number of nodes have a large CP score (close to $1.0$) implying close proximity to the core, while the remaining nodes are in the periphery with a lower CP score. 

The uneven distribution of CP scores suggests a strong Core-Periphery structure in these state networks. Strong Core-Periphery structure is a trait seen generally across all of the state-level networks. The Gini coefficient of CP scores for all nodes in a state is increasing from 2009 to 2014 suggesting that physician involvement in patient care is forming a stronger core-periphery as time passes. 
\vspace{-0.13in}
\subsection{Degree-, clustering-, and connectivity-related statistics}

\noindent \textbf{Degree Distributions and Power Laws.} We computed the in- and outdegree distributions for both the national network and the fifty intrastate networks. The nearly zero p-value of the goodness of fit test against the null hypothesis rejects that the degree distribution is Poisson. Furthermore, the clear difference in terms of clustering coefficient in Table~\ref{table:deg-pl-years} contributes to a rejection of the Erd\'os-Renyi random graph model for the data. 

We next check for a power law. The intuition for considering a power law comes from one of the generative models witnessed in many networks (cf., \cite{MitzPL,NewmanPL}): the so-called ``rich get richer" process. This is perhaps the best known power law-producing generative process, wherein nodes acquire new connections at random but in proportion to their current number of connections. It is plausible that there are groups of physicians (e.g.,  certain types of specialists) that receive and possibly make many more referrals than others and furthermore that physicians accrue new ties in proportion to their existing number of ties. Reputation spread may also manifest as a power law. In contrast, if physicians with many referrals are less likely to accept new referrals  (e.g., they stop taking new patients) and are content with their existing set of ``partner physicians" for referrals, the degree distribution would be expected to be more uniform than depicted by a power law. 

\begin{figure}[h]
  \vspace{-0.14in} 
  \centering
    \includegraphics[height=2.3in,width=2.3in]{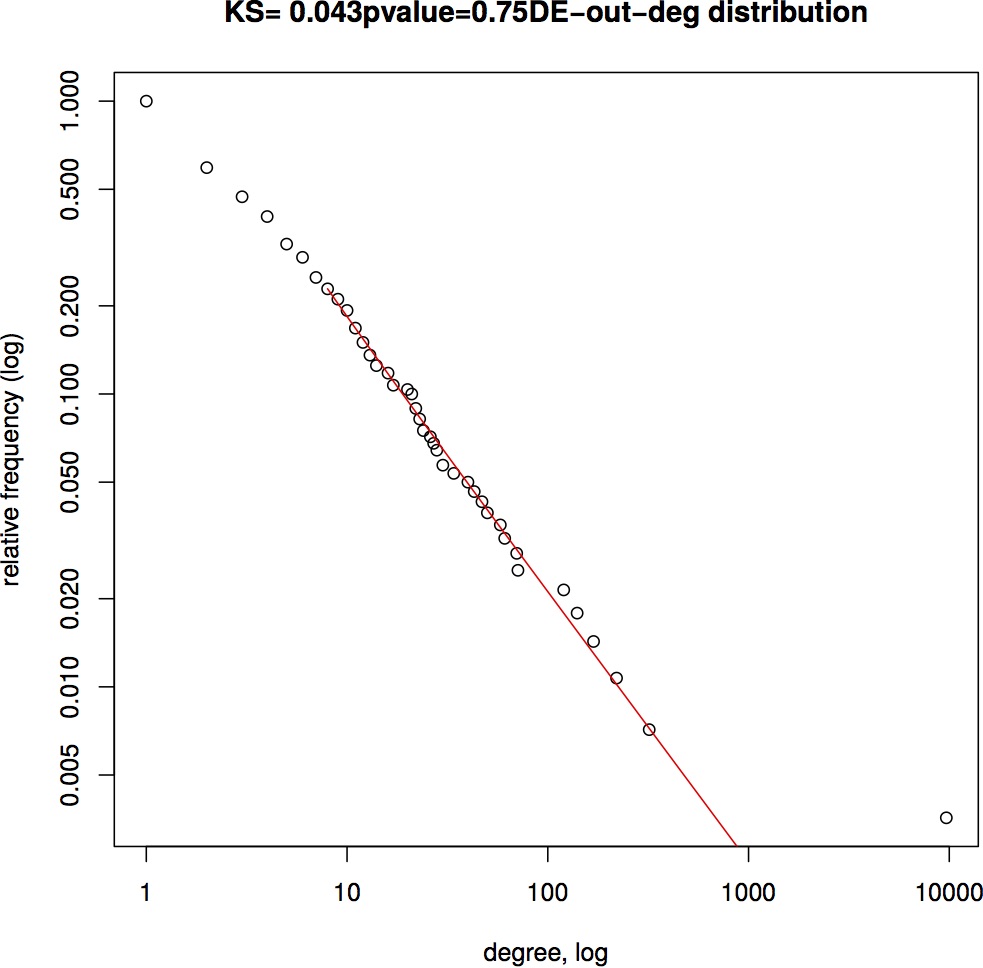}
	\hspace{0.05in}
	\includegraphics[height=1.8in]{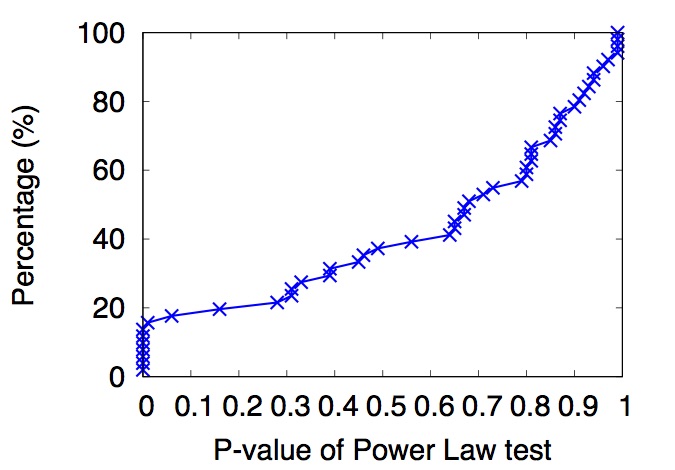}
  \vspace{-0.1in}
  \caption{Out-degree of DE in 2015. P-value distribution of degree-out in 2012 for all states.}
  \vspace{-0.16in}
  \label{fig:deg-pl} 
\end{figure}

\begin{figure}[h]
  \vspace{-0.1in} 
  \centering
    \includegraphics[height=2.0in]{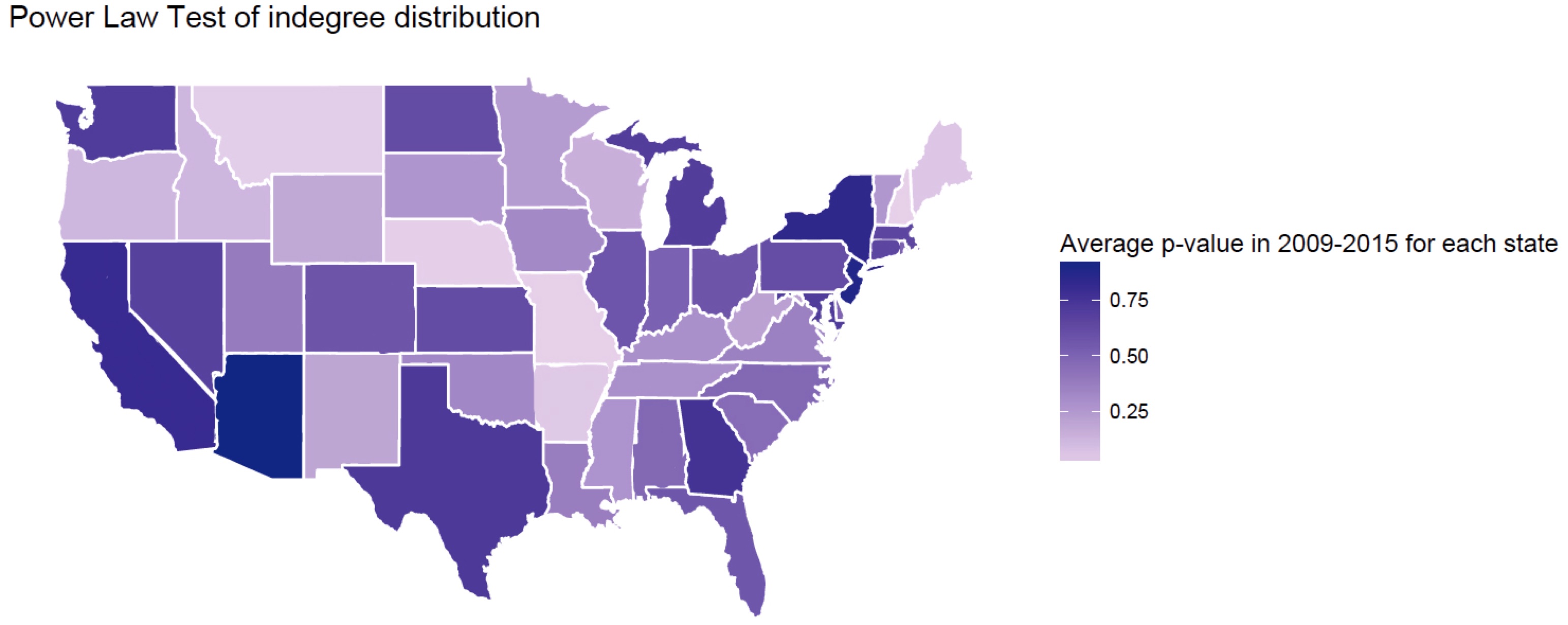}
	\vspace{-0.05in}
	\includegraphics[height=2.0in]{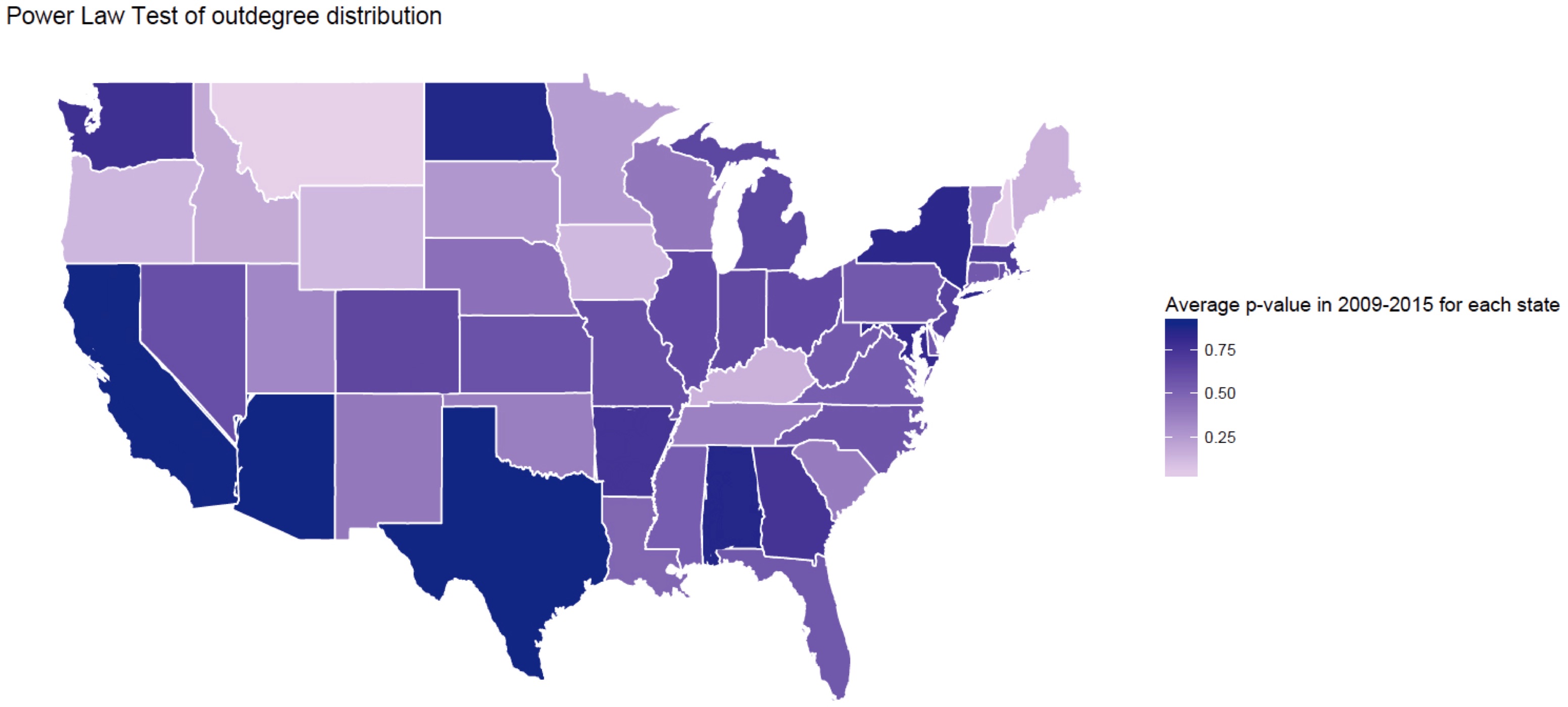}
  \vspace{-0.11in}
  \caption{State map of average p-values of Power Law Test in 2009-2015, for both indegree and outdegree.}
  \vspace{-0.16in}
  \label{fig:deg-pl-maps} 
\end{figure}

In a log-log plot, a power law will appear as a (roughly) straight line. The lefthand of Figure~\ref{fig:deg-pl} shows the power law fitting figure for the 2015 Delaware Intrastate Referral Network. The straight line of the log of Delaware's (unweighted) degree distribution matches the form implied under a power law. The righthand side shows the distribution of the p-value statistic for testing the null hypothesis that the distribution in the network is a power law in the outdegree using the national 2012 data as an example.  We also test the power law hypothesis for both indegree and outdegree among all 50 intrastate networks as well as the whole national network.  For the majority of states (approximately 80\%), the p-values of both the indegree and outdegree distributions are greater than $0.05$, so we fail to reject the null hypothesis that the degree distribution is generated from a power law. Table~\ref{table:deg-pl-years} summarizes the p-value statistics across several years, and 
Figure~\ref{fig:deg-pl-maps} shows the state map of average p-values in 2009-2015. If there was a power law in every state then the p-values for the power law test would follow a uniform distribution. Therefore, to make a claim of whether the power law is universal across all states, we test whether the empirical distribution of the p-values is uniform using the Shapiro-Walks (or a K-S) test. Because the null hypothesis of a uniform distribution is rejected, we conclude that there is strong evidence that the power law is not universal.

The data in Table~\ref{table:deg-pl-years} suggest that the outdegree distributions seem to have a stronger tendency toward power law than indegree. Herein we find the number of states with a p-value $\geq 0.05$. Because a physician does not control who refers patients to them, the number of distinct physicians sending patients may exceed the proportional growth. This is supported by the observation that the indegree distribution has a greater spread than outdegree. Alternatively, the departure of the indegree distribution from a power law might be due to certain specialist physicians being absorbing nodes in the sense that they are the last step in the patient's care (e.g., a sub-specialist). 

\begin{figure}[h]
  \vspace{-0.07in} 
  \centering
    \includegraphics[height=2.0in]{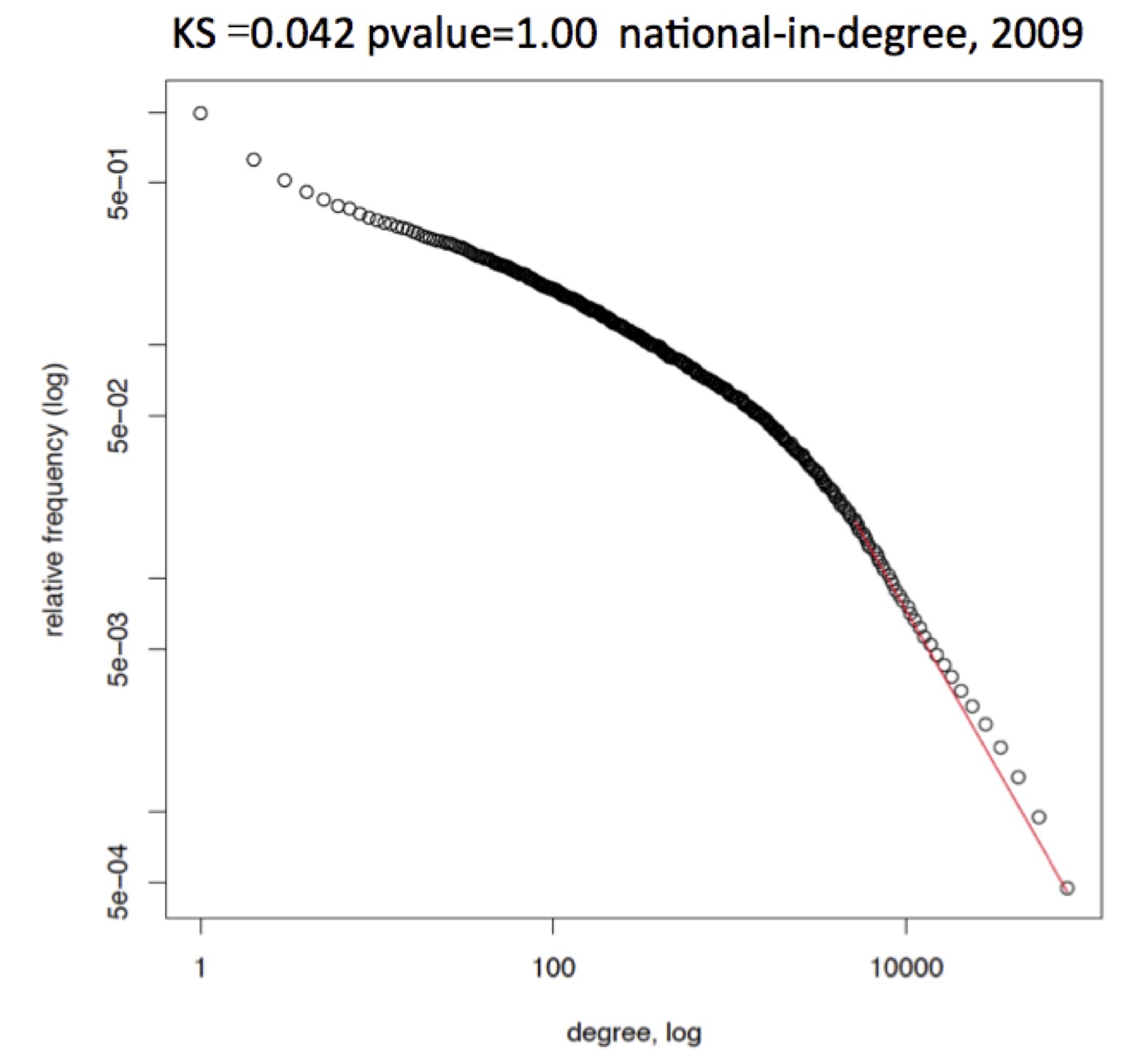}
	\hspace{-0.05in}
	\includegraphics[height=2.0in]{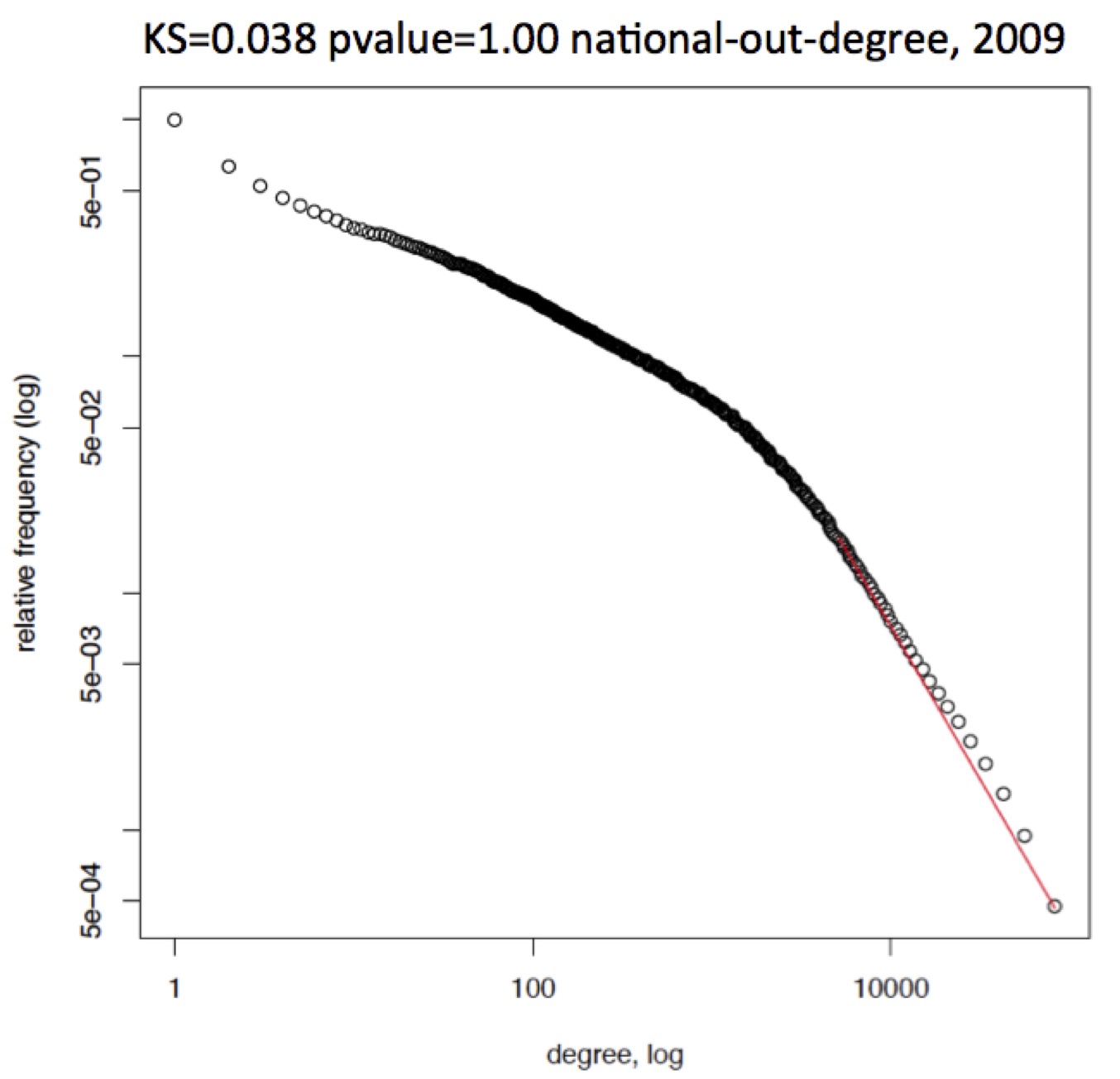}
	\hspace{-0.05in}
	\includegraphics[height=2.0in]{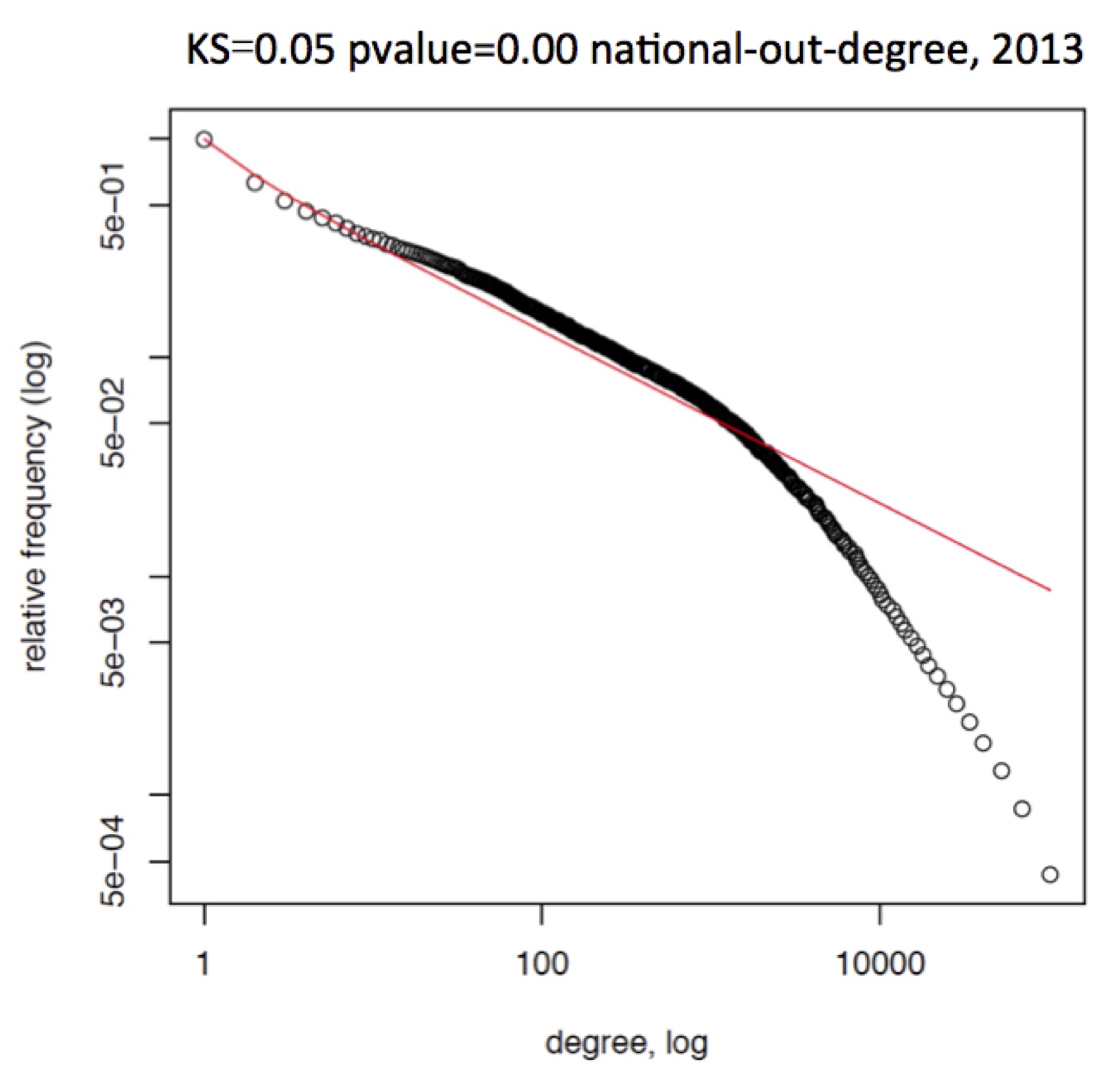}
  \vspace{-0.11in}
  \caption{Degree Power Law of national network.}
  \vspace{-0.16in}
  \label{fig:deg-pl-national-figure} 
\end{figure}

The indegree and outdegree distributions of the national and intrastate networks show power laws with large p-values in Table~\ref{table:deg-pl-years} only except outdegree in 2013. Figure~\ref{fig:deg-pl-national-figure} shows the log-log plot of the three national networks, including the indegree and outdegree log-log plot in 2009 with p-value=$1.0$ and the corner case of outdegree in 2013. The three groups of degree distribution have similar patterns, although the p-values vary.

\noindent \textbf{Assortativity.}
Table~\ref{table:deg-pl-years} displays the average correlation coefficient between two degree values on edges of the 50 state induced referral networks. Given the directed nature of the networks, three kinds of degree assortativity can be measured. We find  (in-,in-) and (out-,out-)  degree correlations exhibit mildly {\bf negative} assortativity, which means patient referral has a small tendency to occur between physicians who possess different levels of indegree or different levels  of outdegree. The significance of the assortativity values against a null hypothesis of no assortativity is tested under an ER null network by using the fact that the asymptotic standard error of $0.5\log((1-r)/(1+r))$ is $SE=(n-3)^{-1/2}=3.35-9.78*10^{-4}$, where $r$ denotes the given Pearson correlation coefficient of the respective degree frequencies and $n$ is the number of physicians in the network. Because the assortativity values are far from 0, it is clear that assortativity is significantly different from 0 in all cases.

\noindent \textbf{Correlation of in-degree and out-degree.}
Table~\ref{table:deg-pl-years} shows the measurement of correlations between indegree and outdegree on the same physician in several years. Since the correlation coefficients in all states are very close to $1.0$, only average values are reported. The results imply that physicians who receive a lot of referrals also make a lot of referrals. The correlation may be inflated due to the fact that specialty is not controlled for and past research~\cite{BarnettML-2012} has shown that degree varies substantially between specialties; if the correlation was measured within physician-type the correlation would likely be lower.

\noindent \textbf{Reciprocity.} If we consider the weight on edges in a directed network, Table~\ref{table:deg-pl-years} shows the R-squared value and correlation coefficient of $w_{ij}$ and $w_{ji}$. The bidirectional weights have strong correlations in different years.
Reciprocity reflects the professional relationship between physicians. The observations support the idea that physicians refer patients back to the referring physician once the specialty appointment is complete or distinct patients see the physician dyad members in opposite orders. Either way, high reciprocity reflects stable collaboration.

\begin{figure}[h]
  \vspace{0.01in} 
  \centering
    \includegraphics[height=1.8in,width=2.7in]{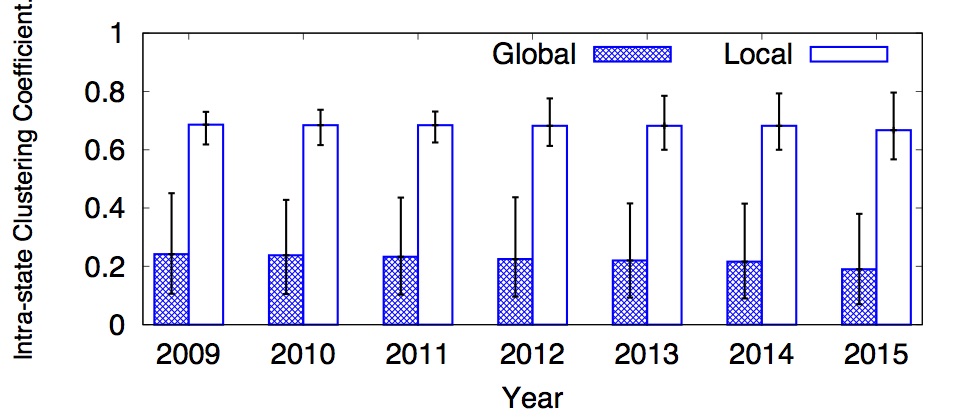}
  \vspace{-0.05in}
  \caption{Clustering coefficients of state network in 2009-2015}
  \vspace{-0.1in}
  \label{fig:cluster-coef} 
\end{figure}

\noindent \textbf{Clustering coefficient.} Figure~\ref{fig:cluster-coef} illustrates both global and local clustering coefficients of states in several years. The error bars show the range of the coefficient values with error bars. 

Table~\ref{table:deg-pl-years} shows the clustering coefficient in the whole national referral network. The local clustering coefficient is much larger than the global one, reflecting a positive correlation between geographic closeness and network flow. The expected local clustering coefficient in an Erdos-Renyi model~\cite{Erdos} $p=\mu/(n-1)$ is much smaller than the measured results. Taken together with the above discussion, we conclude that the patient referral networks have small world character.

\begin{figure}[h]
  \vspace{-0.1in} 
  \centering
    \includegraphics[height=2.0in]{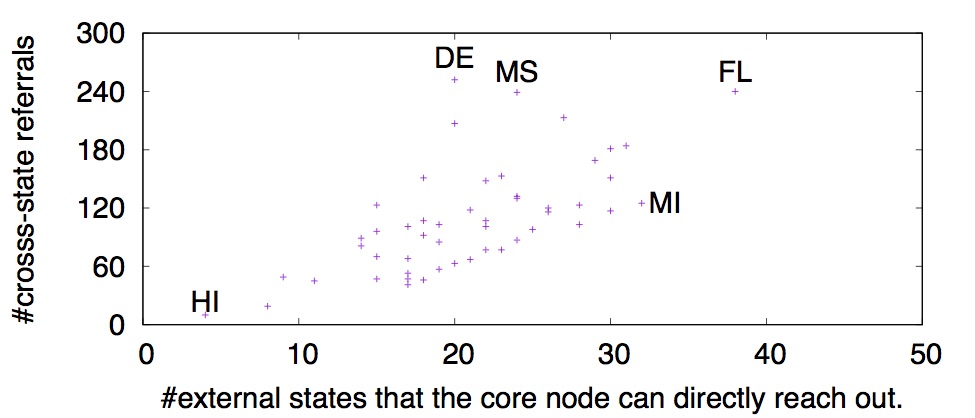}
  \caption{Number of states that a core node with CP score $1.0$ connects to, and the number of  cross-state referrals for the corresponding core node. Time ranges from 2009-2015.}
  \vspace{-0.1in}
  \label{fig:corenodes-state} 
\end{figure}

\noindent {\bf{Core Physician Connectivity Patterns.}}  Each state will have one and only one node with the CP score of $1.0$. We extract those core nodes from each state in 2009-2015 and consider the network constructed by counting the number of cross-state referrals between a core node and external states in Figure~\ref{fig:corenodes-state}.  The core nodes in FL and ML have the widest connectivity, each connecting with physicians in more than 30 other states. On the other hand, the core node in HI only refers or receives patients from 4 external states (presumably a reflection of the distance of Hawaii from the mainland). The core nodes in DE, MS and FL include the most interstate referrals in 2009-2015. The fact that FL has so many referrals is a reflection of the ``snowbird phenomena" whereby a large number of residents of northern states move to southern states, the most popular of which is Florida, for the winter. The same may be partially true of Mississippi.

\begin{figure}[h]
  \centering
    \includegraphics[height=2.3in,width=5.0in]{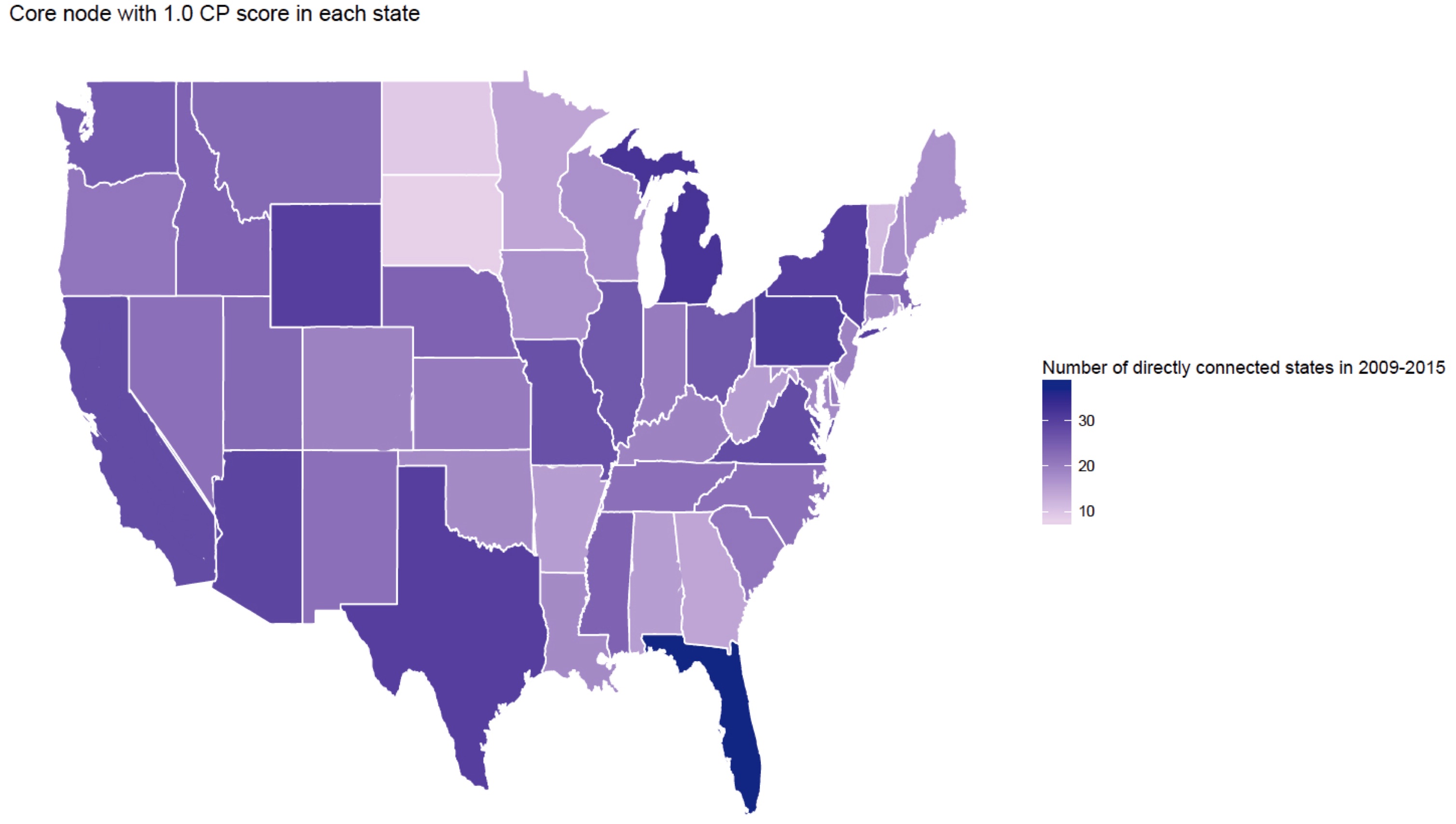}
	\vspace{-0.03in}
	\includegraphics[height=2.3in,width=5.0in]{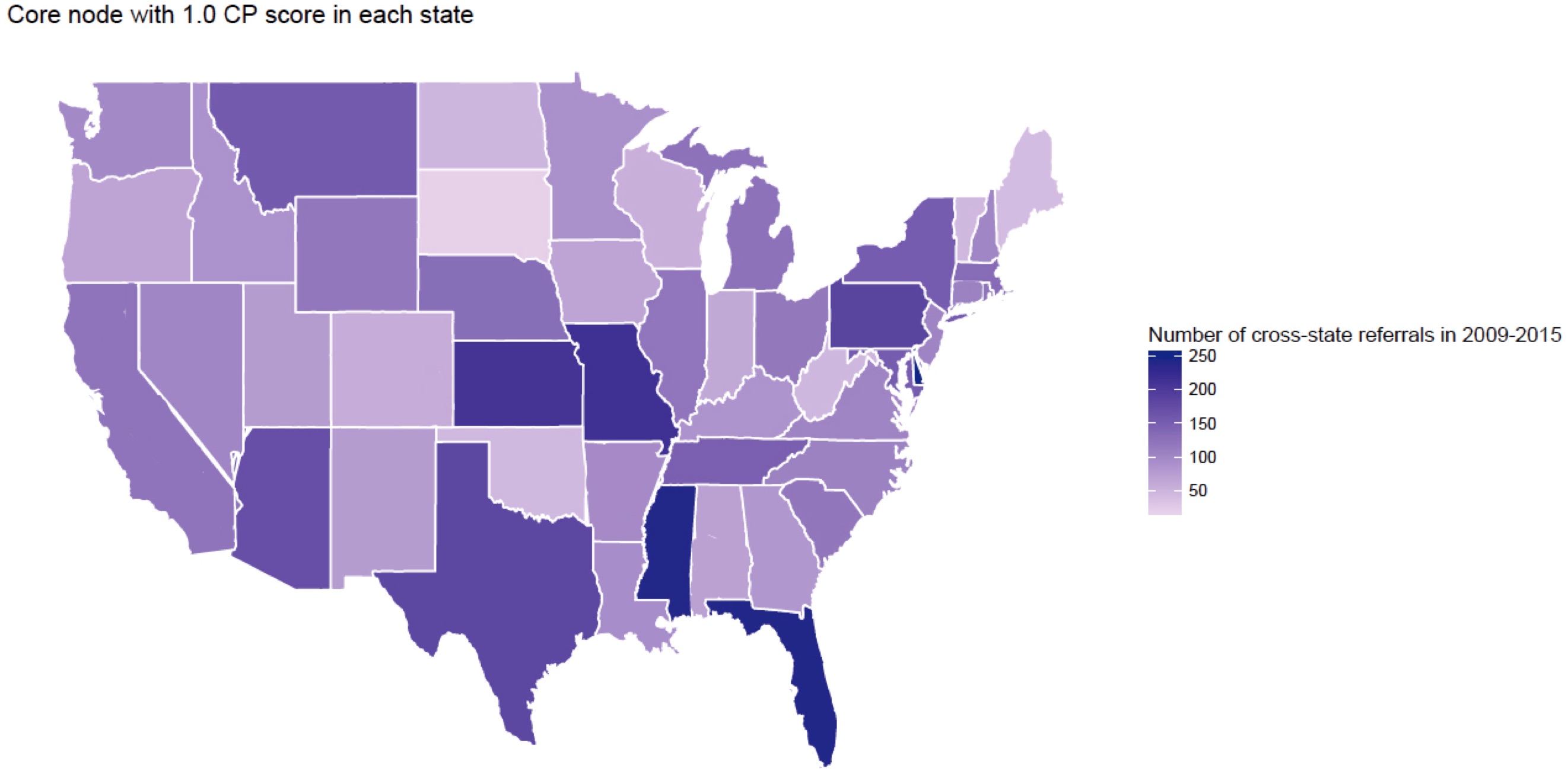}
  \vspace{-0.03in}
  \caption{The state map of Figure~\ref{fig:corenodes-state} and core node connections.}
  \vspace{-0.14in}
  \label{fig:corenodes-maps} 
\end{figure}

\begin{figure}[h]
  \vspace{-0.04in} 
  \centering
    \includegraphics[height=2.3in]{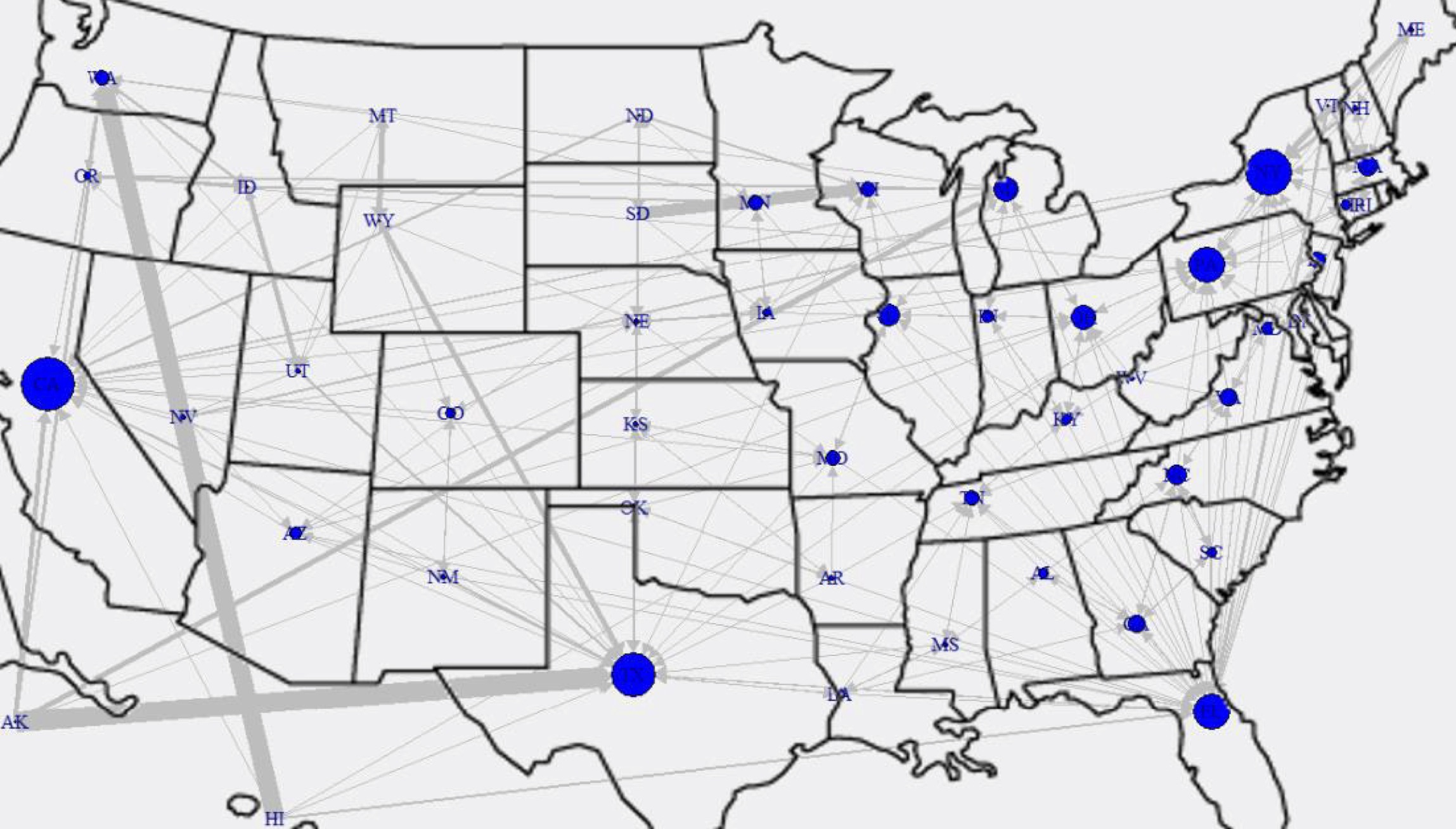}
  \caption{A visualization of the top five out-of-state referrals. Edge width varies by percentage of within-state referrals, and node size varies by size of state's physician referral network. The map depicts directed edges of each state's top five states to which they refer patients. Note that nodes for Alaska (AK) and Hawaii (HI) are placed in the bottom left of the figure. As this is a network diagram, the placement position is not relevant.}
  \vspace{-0.13in}
  \label{fig:topstates} 
\end{figure}

\noindent{{\bf{A Skeletal National Network and Gravity Law.}}} Figure~\ref{fig:topstates} displays for each state the connections to the top five states to which its physicians make referrals. NY, PA, FL, TX, CA, MI, MA play the role of ``hub'' states, while  midwestern states do not have many cross-state referrals. In general, the traffic map matches the population distribution in the U.S. 

The flow of patients around the country -- as represented by referral data -- suggests that the tools and models of economic geography might be relevant. Most famous is the well-known ``gravity law" of the type defined by Equation~\ref{equal:gravity-law}: 
\begin{equation}
\label{equal:gravity-law}
F_{ij} = G\frac{M_{i}^{\beta_{i}}M_{j}^{\beta_{j}}}{D_{ij}^{\beta_{d}}},
\end{equation}
where $F_{ij}$ represents the number of distinct patients referred from state $i$ to state $j$, $G$ is a constant, $M_{i}$ and $M_{j}$ represent the number of physicians in states $i$ and $j$, respectively, and $D_{ij}$ denotes the distance between the two capital cities. Such a ``law" is commonly fit to macroeconomic trade patterns. \cite{PRKrugman1997} Based on referral records in 2009-2014, an OLS method is used to estimate the $\beta$ coefficients. $\hat{\beta}_{i}=0.904, \hat{\beta}_{j}=0.904, \hat{\beta}_{d}=1.342$ so that the number of distinct patients is inversely proportional to the distance, which validates the hypothesis of a gravity law. As further support, in a log-log plot the linear relationship between $F_{ij}$ and $M_{i}, M_{j}, D_{ij}$ is significant with p-value $<2.2e-16$, residual standard error 2.258, R-squared 0.5764. 

\subsection{Motif analysis} 
\begin{figure}[h]
	\center
	\vspace{-0.02in}
\includegraphics[width=10cm,height=6.6cm]{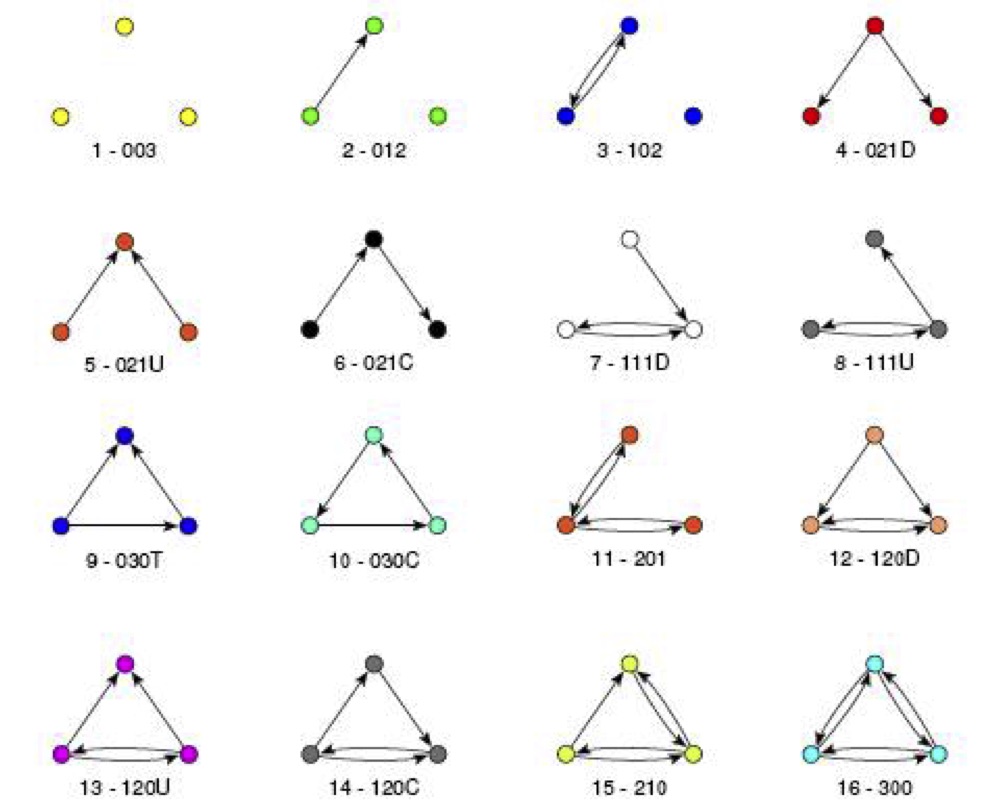}
   \vspace{-0.01in}
   \caption{\textbf{16 kinds of triads}}
   \label{fig:Result-16triads}
   \vspace{-0.15in}
\end{figure}

Network ``motifs" are commonly recurring small patterns of connectivity, often thought of as a network's ``building blocks". \cite{MiloMotifs} Dyadic motifs are the simplest in structure having just two nodes and in a directed binary-valued network only a few possible states. If the motifs do not distinguish between the edge from physician A to B and that from B to A, there are only three dyadic paterns: no edge, one directional edge and bidirectional edges. While no-edge case is dominant in terms of frequency, the fraction of Monte-Carlo estimated frequency of directional dyads and bidirectional dyads is around 24:76, implying a very high-level of reciprocity is present in the network. 
As a part of the exploration of patterns in patient referral networks we engaged in exploratory analysis to discover what kinds of triads in our directed networks are most prevalent. Figure~\ref{fig:Result-16triads} illustrates the 16 possible triads. 

Table~\ref{table:global-triad} displays the Monte-Carlo estimated frequency of the various triad structures (i.e., randomly sample node 3-tuples and record the connectivity structure) over 2009-2015 in the national network (a Monte Carlo calculation of $10^8$ random draws was used because complete enumeration is infeasible). The completely disconnected triad (Triad 1 in Figure~\ref{fig:Result-16triads}) is far and away the most prevalent and we do not record its number.  The remaining 15 kinds of triads break up naturally in terms of order of magnitude of frequency into 7 groups: (1) Triads 2 and 3: Two physicians share patients in one or two directions; (2) Triad 11: a physician shares patients with two physicians mutually; (3) Triads 7 and 8: a physician shares patients with one physician mutually and with another physician in only one direction; (4) Triads 5,6,15, and 16: loose connections and close connections between three physicians; (5) Triads 12, 13, 14: a pair of mutually connected physicians with the third physician whose degree is two; (6) Triad 9: a triple that follows transitivity (if A refers a patient to B and B refers a patient to C the chance that A referred a patient to C is substantially greater than otherwise) and lastly (7) Triad 10 without transitivity. 

Since the referral records do not contain patient ID, we cannot track the same patient and analyze the referral sequence. The rank order remains roughly the same over each year suggesting that the structure of the network is stable in this regard. Triad 2 and Triad 3 are the two most popular triad patterns in the whole referral network, accounting for the majority of the triads in the state networks. These convey two of the most elementary care patterns. Under Triad 2, a patient encounters physician A followed by physician B and then is done. Under Triad 3 the patient emulates the care pattern of Triad 2 but then returns to see physician A again. The frequency distribution suggests that the network contains regions of high density, or even cliques, since some triads with more edges (T15 and T16), representing more complex care patterns within reciprocated referrals between 2 or 3 physicians, occur more frequently than triads with fewer edges (T9, T10, T12, T13, T14). 

To detect the latent groups of triads, we remove the columns of the correlation matrix corresponding to Triads 2 and 3, which remain as their own groups, and perform a factor analysis on the correlation matrix for Triad types 4-16. Table~\ref{table:factor} displays the factor loadings on those triads for 2009-2015. Assuming two factors, the groups of triads based on highest loading correlation are (T4, T9, T10, T12, T13, T14, T15, T16) and (T5, T6, T7, T8, T11). With three latent factors, the loadings also support this group division as for the two-factor case. 

The factor analysis provides more flexibility for the regression analyses performed in Section~\ref{sec:healthcare}. For these analyses we have limited observations, 50 states observed in 5 years for a total of 2,500 observations. The clustering of observations within states reduces the information content of the data. Our ideal model would simultaneously estimate the effect of all network-based predictors, and any other predictors, on each of these dependent variables. With a modest number of observations, the model is vulnerable to over-specification if an excessive number of predictors are included. As opposed to including 15 measures of the relative frequency of each type of triad, we instead may include the relative frequency measures for two of the three groups of triads allowing more flexibility in regards to including other predictors, interaction variables, and transformed predictors. In building the hierarchical model in Section~\ref{sec:healthcare}, we test whether replacing the individual triad predictors with their factor-based groups was justified by comparing the total variation explained by all of the triad indicators to that explained by the grouped variables and testing the significance of the difference. To ensure that the grouped triadic variables captured triadic or higher-level effects and not dyad-level effects, the level of reciprocity in each state and year was also included as a predictor in any model containing triad-level predictors.

\subsection{Diversity among states} 
From hereon, we discard the data in 2015 since the period of observation is not complete and many healthcare attributes are not available in 2015. 
\begin{figure}[h]
  \vspace{-0.5in} 
  \centering
    \includegraphics[height=3.3in]{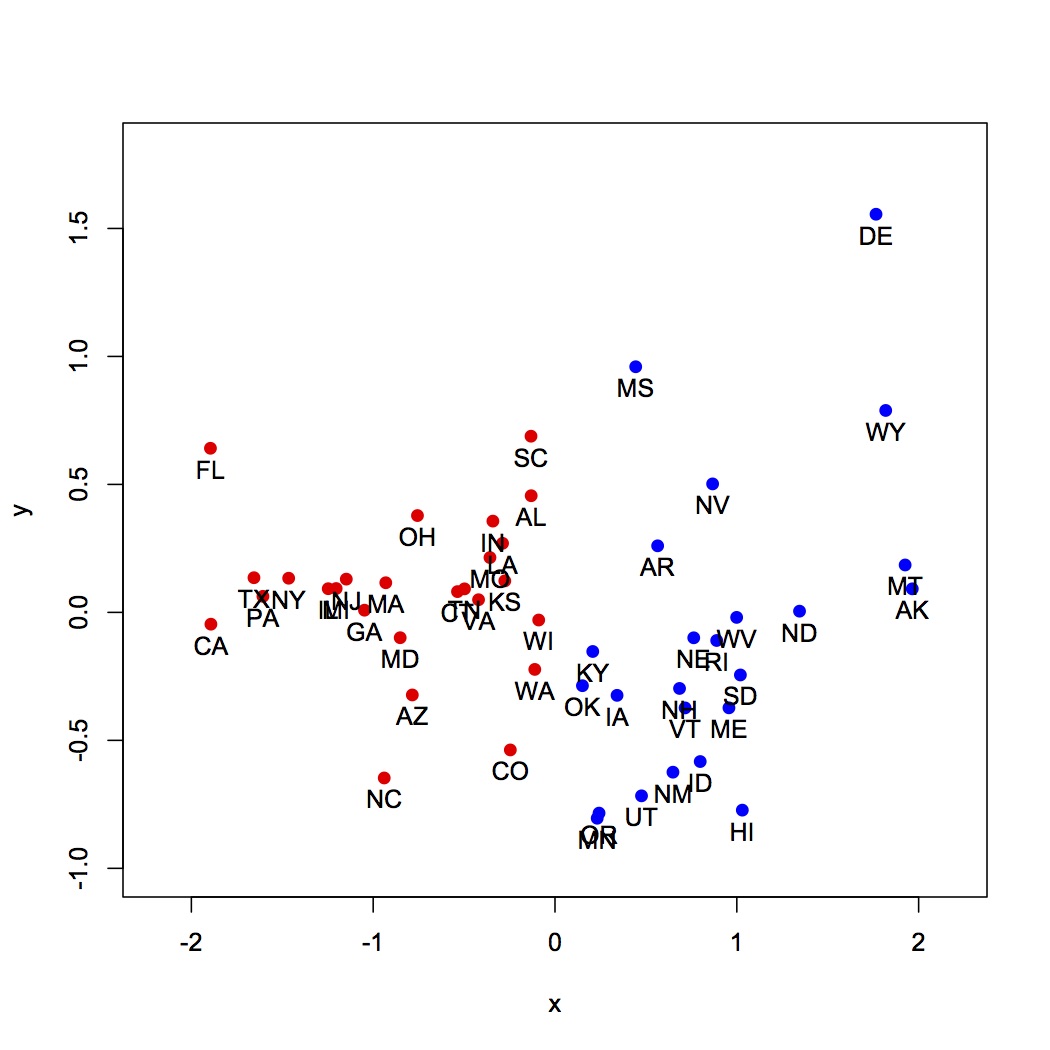}
	\vspace{-0.15in}
  \caption{Multidimensional scaling (MDS) plot of 50 states based on feature vector in 2014. Two clusters are in red and blue. }
  \vspace{-0.15in}
  \label{fig:MDS-plot} 
\end{figure} 
We apply the K-means clustering algorithm to the 50  feature vectors defined by the state-level network measures in Table~\ref{Appendix-predictors}. Figure~\ref{fig:MDS-plot} is a 2-d visualization, produced via multidimensional scaling (MDS). The red and blue coloring of the nodes represents the outcome of applying K-means with two clusters, for which the centroids are MA (red) and ME (blue) respectively. We find that the cluster represented by MA generally includes the states which   have more physicians or larger population than those of the cluster represented by MA.  Table~\ref{table:cluster-diverstiy} shows the centroid states of each cluster for $K = 2, 3, 4, 5$.
\vspace{-0.12in}
\section{Comparative Analysis: Relationships Between Network and Healthcare Measures} 
\label{sec:healthcare}

Table~\ref{table:network-state-connection} shows selected univariate correlations (i.e., not accounting for the possible overlapping effects of any other predictors) between network measurements and healthcare statistics in 2009-2014 for all state-years. The presence of such high correlations suggest the potential for estimating a multiple regression model that exhibits a high level of predictive accuracy. Because the data contain repeated measurements on states, our core statistical regression model is a mixed-effect model with state random effects. Denoting an outcome and vector of network measures for state $i=1,\ldots,50$ in year $t=0,\ldots,5$ by $Y_{it}$ and $X_{it}$, respectively, the mathematical specification of these models has the following general form:
\begin{equation}
\label{equal:mlr-vary-intercept}
Y_{it} = \alpha_{i}+\lambda_{t} + \bm{\beta}_{1}^{T} \bm{X}_{it} + \bm{\beta}_{2}^{T} \bm{X}_{it}t + \varepsilon _{it}
\end{equation}

where $\alpha_{i} \sim N(\beta_{0},\tau^{2})$ describes the distribution of the state-level random effects and $\varepsilon _{it} \sim N(0,\sigma _{y}^2)$ describes the distribution of the error term. The main effect of the vector of network measures is $\beta_{1}$ while its modification by linear year ($t$) is $\beta_{2}$. The parameters $\lambda_{t}$ allow for an unstructured population trend across years whereas the state-specific departures from this trend are restricted to be linear (i.e., proportional to $t$). Modeling the time-trend as an unstructured (maximally flexible) main effect and linearly as a network effect modifier serves the purpose of fully capturing the population trend (the variation we want to control for) and parsimoniously representing the effect-modification by time (the effect of interest that we want to interpret). It also allowed for more stable estimation than when seeking to estimate separate effects of the network measures in each year.

Separate models of the form given in Equation \ref{equal:mlr-vary-intercept} were estimated for each of the 20 state-level health attributes. Therefore, the initial model for each dependent variable allowed the coefficient of the network predictors to change linearly with time. If the interaction of linear time with a given network measure is non-significant, it was dropped from the final specification. To determine the model for each health care outcome, we used a model building process that sought to obtain a parsimonious set of models with a common set of predictor variables across the outcomes. Starting with the predictor with the strongest univariate effect, we added predictors one at a time until no more predictors that had significant effects could be added. We repeated the same process across all outcomes. The final set of network measures was obtained by taking the union of all included network measures across the outcomes; we denote the size of this set by $M$. For each outcome, we then added all of the interactions of these predictors with $t$ and starting with the least significant dropped them one-by-one until only significant terms remained. For both the inclusion and exclusion steps, we use a chi-squared test of the magnitude of variation in the outcome across the levels of the predictor to determine if the predictor accounts for a significant amount of variation, justifying its inclusion in the model, or if it adds minimal or no explanatory power and can be excluded. The final models thus contained the same network measures but differed in terms of how of them many had significant linear interactions with time. For example, when medical-discharge-per-1000-people (called f34) is the dependent variable, none of the selected network features had a significant linear interaction with time, so the interaction predictor is absent. 
 
As Table~\ref{tab:MLR-more-results} reveals, the selected features for the regression of medical-discharge-per-1000-people (f34) are: average-degree (f1), (out-, out-) assortativity  in induced-network (f25) and the Gini coefficient of sizes of the components (groups of nodes with no edges between them) of the induced network (f16). Table~\ref{table:coefficient-fixed-effects} shows the change of coefficients for selected predictors from model 1 to model 3 in Table~\ref{tab:MLR-more-results}. The coefficients range from 4.08 to 14.70 and are directly comparable as the predictors were standardized to a mean of 0 and standard deviation of 1. According to m3, the dependent variable (f34) is more sensitive to f1, and changes in the referral network are associated with healthcare attributes.
\vspace{-0.08in} 
\subsection{Result analysis}
Some features (e.g., f1, f5, f9, marked with asterisk) have significant interactions with time (each year) for several healthcare attributes in Table~\ref{tab:MLR-more-results}, because they are related to the increasing size of the referral network (Table~\ref{table:ref-data}). The coefficients of interaction items are in Table~\ref{Appendix-interaction-coefficients}. None of the triadic predictors warranted inclusion in the final model for each outcome.

Across individual healthcare measures the most prominent network-based predictors involve the physician degree (the density of the network -- f1), number of physicians in the intra-state network (f9), out-degree (out-,out-) assortativity of the induced network, and heterogeneity of the size of the networks components as computed by the Gini coefficient (f16). The general assortativity results imply selectiveness of physicians to seek out physicians like themselves (i.e., those with similar patient referral tendencies) when making patient referrals while the component size distribution reflects co-existence of large provider groups and independent practices in a region. It is unclear whether these are characteristics of a region that could be exploited to incite more efficiency or otherwise improve healthcare. For further illustration, we consider significant network predictor-outcome pairs and provide illustrative interpretations with respect to particular outcomes, recognizing that these points only suggest a direction of possible investigation, but lack  obvious explanations.

\begin{itemize}

\item For medical-discharge-per-1000-people the high associations are average degree (f1), assortativity (f25) and Gini coefficient of component size (f16). The positive coefficients might reflect that more discharges from hospital  might lead to or generate more follow-up care. 
\vspace{-0.025in}
\item Hospital-admissions-per-1000-decedents is understandably highly correlated with the number of patient referrals. This might be a front-end of the above result -- both hospitalizations and discharges from hospital may be a result of or generate more referrals.
\vspace{-0.025in}
\item For long-term-opioids-receipt (see \cite{Meara2016} for description), the clustering coefficient (f7 and f20) has a significant negative effect suggesting that networks with greater closure better contain overall opioid consumption (e.g., through better patient monitoring). 
\vspace{-0.025in} 
\item Gross State Product (GSP) is associated with the Gini coefficient of in-degree; one possible explanation is that having a few physicians who receive an extreme number of referrals might be a marker of a state having the capability to perform rare, highly complex procedures.
\vspace{-0.025in} 
\item For the regression of \#nodes-per-1000-people both average degree and global clustering coefficient have large estimated coefficients. A larger average degree implies that more physicians are working in the state, possibly indicating a form of supply driven utilization.~\cite{Oosterhaven1988}
\vspace{-0.025in}
\item A uniform distribution of physicians and a low clustering coefficient are associated with more physicians being shared per 1,000 people. This may indicate that patients do more physician shopping (searching for physicians who'll approve the pharmaceuticals or procedure they seek) in states whose physician networks have less closure (i.e., less clustering). 
\vspace{-0.025in} 
\item Mortality is associated with average degree and \#edges. This is an interesting finding but before making conclusive interpretations it would be important to control for patient demographic and clinical variables.
\vspace{-0.025in}
\item Gini coefficient of out-degree (f5) has a large association with expense-per-inpatient-day suggesting that patients on average pay more for treatment in states with greater heterogeneity of physician popularity. Again, this may be a marker for the availability of specialty services in the state. 
\vspace{-0.025in}  
\item For the regression of \#hospital-beds-per-1000-people the Gini coefficient measures (f5 -- out-degree distribution and f16 -- component size) both have negative estimated coefficients suggesting that the more that care is regionalized at select institutions (perhaps by those performing the majority of  a given specialty or performing some kind of other non-standard care), the less the need for hospital beds.
\end{itemize}
\vspace{-0.2in}
\section{Conclusion}
\label{sec:conclusion}
In this paper, we applied algorithms and methods from  graph theory, network science and statistics to explore many network features in the U.S. patient referral networks. Those network features describe both micro and macro patterns about patient referrals, such as power laws in some degree distributions, ``small world'' structure, Core-Periphery structure, motifs of triadic structures, and ``Gravity Law" in cross-state referrals. Indeed, with the last of these we  suggest implicit broader connections with economic geography (see e.g., \cite{PRKrugman1997}) which are intriguing and suggest natural follow-ons. We also find close correlations between network features and state healthcare attributes. A better understanding of network features may provide insights into directions for improving the healthcare system. Our findings come with the caveat implicit that the data only includes referrals within the Medicare system. Nevertheless, the fact that we use only data that is available to the public means that our methods and results can be replicated. This furthermore supports one of our key goals of introducing the growing network science toolkit (measures, tools, and models) to statisticians and medical researchers who are less familiar with generative models in particularly as well as empirical models for social network data, and to also show that these ideas can be deployed on a very large scale. 

From 2009-2015 we found that the majority of network features are fairly stable. Our key results encompass both general or macro-level and micro-level network features. At the macro-level, the power law structure cannot be rejected in most cases, which suggests that these networks are ``robust yet fragile" --  i.e., robust to random failure, but susceptible to ``targeted"  attack (i.e., consciously specified removal). \cite{AlbertNature} The small-world property implies that physician networks are suitable for efficient information transfer and diffusion of innovations. \cite{WattsSWP} Analyses at both state and national network level tends to support the hypothesis of a ``small world'' and thus a fertile environment for diffusion (see also \cite{StrogatzNature} and \cite{Koss} for other possible connections) and suggests a rich direction of future research. At a micro-level, the computation of actor specific network measures allows rankings of physicians to be constructed based on their importance in the referral network. Possible measures that can be used include degree, local clustering coefficient, CP score, and the number of external connections. 

By evaluating associations between the network measures for each state and health variable, we gain insights into how improvements in health care organization and ultimately outcomes might be incentivized. For example, we found that clustering was positively associated with per-capita state opioid prescription fills. Therefore, to reduce over-use of opioids, a public health intervention that sought to limit referrals among the highest prescribing physicians could be investigated.

In general, to optimize the referral network in terms of healthcare system efficiency and physician workload, we hope that lessons can be learned from understanding the network aspects of the most well-performing states and consequent investigation of whether it might be useful to try to extend or even enforce such structures  in other states. A natural next study that still only uses freely available data would be to incorporate physician specialty (unlike patient data, information on physicians is publicly accessible) in order to both distinguish physician network effects from physician specialty effects and also to reveal specialty-network interaction effects. However, the ultimate goal is to link the national physician network to individual patient data in order to perform patient-level analyses that account for patient demographic and clinical factors when assessing the association of the physician network and its salient subnetworks to important patient outcomes.

\bibliographystyle{abbrv}
{\footnotesize
\bibliography{reference}}

\newpage
\begin{table}[h!]
\footnotesize
\centering
\vspace{0.2in}
\caption{Dataset size by year.}
\begin{tabular}{|c c c c c c c c|} 
 \hline
 Year & 2009 & 2010 & 2011 & 2012 & 2013 & 2014 & 2015 \\ [0.5ex] 
 \hline
 \#Records & 50,382,951 & 52,236,906 & 54,038,549 & 54,966,715 & 55,141,669 & 55,779,360 & 34,856,901 \\ 
 \hline
 \#Physicians  & 890,452 & 921,959 &955,659& 987,770 & 1,018,245 &1,044,758& 961,309  \\ \hline
\end{tabular}
\label{table:ref-data}
\end{table}

\begin{table}[h]
\scriptsize
\centering
\vspace{-0.25in}
\caption{P-values for rejecting various Power Laws, assortativity, self-degree-correlation, reciprocity and clustering coefficient of the national patient referral network (or average among states) in 2009-2015.}
\label{table:deg-pl-years}
\begin{tabular}{|l|l|l|l|l|l|l|l|}
\hline
Year                                         & 2009   & 2010   & 2011                    & 2012   & 2013   & 2014   & 2015   \\ \hline
in-degree p-value of national network Power Law     & 1.00   & 0.99   & 1.00                    & 1.00   & 0.93   & 0.91   & 0.38   \\ \hline
\#states in-degree p-value\textgreater0.05  & 32     & 37     & 37 & 36     & 36     & 37     & 36     \\ \hline
Average p-value of in-degree Power Law among states               & 0.4084 & 0.4084 & 0.4137                  & 0.4245 & 0.4423 & 0.4388 & 0.4654 \\ \hline
out-degree p-value of national network Power Law    & 1.00   & 1.00   & 0.99                    & 0.97   & 0.00   & 0.97      & 0.74   \\ \hline
\#states out-degree p-value\textgreater0.05 & 39     & 43     & 42                      & 37     & 37     & 38     & 40     \\ \hline
Average p-value of out-degree Power Law among states              & 0.4545 & 0.5292 & 0.5913                  & 0.5303 & 0.5190 & 0.4956 & 0.4484 \\ \hline
Average (in, in) assortativity among states   & -0.1084 & -0.1083 & \multicolumn{1}{r|}{-0.1101} & -0.1126 & -0.1132 & -0.1137 & -0.1217 \\ \hline
Average (out, out) assortativity among states & -0.1104 & -0.1108 & -0.1125                      & -0.1150 & -0.1157 & -0.1161 & -0.1245 \\ \hline
Average (in, out) assortativity among states  & 0.0775  & 0.0752  & 0.0727                       & 0.0692  & 0.0662  & 0.0633  & 0.0549  \\ \hline
Average (out, in) assortativity among states  & 0.0800  & 0.0775  & 0.0750                       & 0.0714  & 0.0684  & 0.0654  & 0.0569  \\ \hline
State self in/out degree: average R-squared value         & 0.9717 & 0.9715 & 0.9712 & 0.9717 & 0.9710 & 0.9711 & 0.9692 \\ \hline
State self in/out degree: average correlation coefficient & 0.9858 & 0.9856 & 0.9855                      & 0.9857 & 0.9853 & 0.9854 & 0.9845 \\ \hline
State reciprocity: average R-squared value         & 0.9074 & 0.9094 & \multicolumn{1}{r|}{0.9073} & 0.9053 & 0.9045 & 0.9015 & 0.8927 \\ \hline
State reciprocity: average correlation coefficient & 0.9524 & 0.9535 & 0.9524                      & 0.9513 & 0.9509 & 0.9493 & 0.9445 \\ \hline
global clustering coefficient of national network & 0.0763  & 0.0740  & \multicolumn{1}{r|}{0.0727} & 0.0682  & 0.0623  & 0.0609  & 0.0523  \\ \hline
local clustering coefficient of national network & 0.700   & 0.699   & 0.698                       & 0.698   & 0.698   & 0.699   & 0.691   \\ \hline
E(C) by Erd\'os-Renyi Model of national network     & 1.27e-4 & 1.23e-4 & 1.18e-4                     & 1.13e-4 & 1.06e-4 & 1.02e-4 & 7.54e-5 \\ \hline

\end{tabular}
\end{table}

\begin{table}[ht]
\centering
\vspace{-0.25in}
\caption{Triad frequencies for the U.S. national referral network.}
\label{table:global-triad}
\centering
\resizebox{\textwidth}{!}{\begin{tabular}{|l|l|l|l|l|l|l|l|l|l|l|l|l|l|l|l|}
\hline
ID & 2        & 3        & 4   & 5    & 6    & 7     & 8     & 9  & 10 & 11     & 12  & 13  & 14  & 15   & 16   \\ \hline
2009          & 23433902 & 76245745 & 188 & 4096 & 5113 & 56061 & 28650 & 66 & 1 & 222166 & 171 & 127 & 157 & 1484 & 2073 \\ \hline
2010          & 23747795 & 75929710 & 206 & 4426 & 5321 & 58748 & 28634 & 49 & 4 & 221342 & 176 & 117 & 141 & 1427 & 1904 \\ \hline
2011          & 23865204 & 75802642 & 175 & 4999 & 5720 & 62887 & 29448 & 56 & 0 & 225125 & 166 & 124 & 144 & 1342 & 1968 \\ \hline
2012          & 23892310 & 75764994 & 167 & 4989 & 5811 & 64475 & 30656 & 63 & 1 & 233164 & 163 & 99  & 127 & 1264 & 1717 \\ \hline
2013          & 24202517 & 75439104 & 180 & 5648 & 6524 & 68943 & 31777 & 51 & 5 & 242030 & 155 & 114 & 125 & 1185 & 1642 \\ \hline
2014          & 24405405 & 75233266 & 201 & 5803 & 6571 & 69167 & 32792 & 51 & 4 & 243538 & 174 & 104 & 109 & 1210 & 1605 \\ \hline
2015          & 25421893 & 74265148 & 147 & 5160 & 6326 & 59787 & 31480 & 52 & 2 & 207622 & 140 & 90  & 86  & 948  & 1119 \\ \hline
\end{tabular}}
\end{table}
\begin{table}[ht]
\centering
\vspace{-0.02in}
\caption{Factor analysis loadings of state level triad terms over 2009-2015.}
\label{table:factor}
\centering
\resizebox{\textwidth}{!}{\begin{tabular}{|l|l|l|l|l|l|l|l|l|l|l|l|l|l|}
\hline
\#Factors=2   & T4    & T5    & T6    & T7    & T8    & T9    & T10    & T11   & T12   & T13   & T14   & T15   & T16    \\ \hline
Factor1       & 0.853 & 0.608 & 0.536 & 0.348 & 0.346 & 0.872 & 0.909 & 0.182 & 0.896 & 0.912 & 0.923 & 0.926 & 0.938  \\ \hline
Factor2       & 0.513  & 0.756 & 0.836 & 0.880 & 0.868 & 0.453 & 0.399  & 0.687 & 0.408 & 0.397 & 0.397  & 0.357 & 0.278  \\ \hline
\#Factors=3 & T4    & T5    & T6   & T7    & T8    & T9    & T10    & T11   & T12   & T13   & T14  & T15   & T16    \\ \hline
Factor1       & 0.870 & 0.644 & 0.574 & 0.370 & 0.372 & 0.894 & 0.920 & 0.175 & 0.920 & 0.929 & 0.939 & 0.937  & 0.944  \\ \hline
Factor2       & 0.466 & 0.654 & 0.753 & 0.897 & 0.876 & 0.358 & 0.367 & 0.815 & 0.309 & 0.342  & 0.326 & 0.346 & 0.297   \\ \hline
Factor3       & 0.123 & 0.329 & 0.279 & 0.110 & 0     & 0.249 & 0     & -0.116 & 0.224 & 0.109 & 0     & 0     & -0.129 \\ \hline
\end{tabular}}
\end{table}
\newpage
\begin{table}[]
\centering
\vspace{0.3in}
\caption{The nearest states to the centroids of the clusters of states obtaining using K-means.}
\label{table:cluster-diverstiy}
\begin{tabular}{|l|l|l|l|l|}
\hline
Year\textbackslash \#cluster & n=2   & n=3      & n=4         & n=5            \\ \hline
2009                         & ME MA & LA NC SD & NC OR SD TX & KY MA OR SD TX \\ \hline
2010                         & ME NC & LA PA SD & LA OR PA SD & LA ME OR PA SD \\ \hline
2011                         & ME MD & IL PA SD & LA NM PA SD & LA MT NM PA SD \\ \hline
2012                         & ME MA & ME PA TN & ME PA SD TN & LA ME MT PA TN \\ \hline
2013                         & ME MA & ME MI TN & MI MT OR TN & MI MT NE OR TN \\ \hline
2014                         & ME MA & ME MI MT & ME MI MT TN & ME MI MT NE TN \\ \hline
\end{tabular}
\end{table}

\begin{table}[]
\centering
\vspace{0.2in}
\caption{Selected correlations between network measures and state healthcare measures.}
\vspace{0.03in}
\label{table:network-state-connection}
\begin{tabular}{|l|l|l|l|}
\hline
Network                      & Healthcare                   & R-square & Correlation coefficient \\ \hline
\#edges                      & \#population                 & 0.813    & 0.902                   \\ \hline
average degree               & medical reimbursement        & 0.662    & 0.814                   \\ \hline
Gini coefficient of CP score & \#physicians                 & 0.578    & 0.760                   \\ \hline
assortativity (in, in)       & \#medicare beneficiaries     & 0.491    & 0.701                   \\ \hline
assortativity (in, out)      & state expenditure per person & 0.317    & 0.563                   \\ \hline
average degree               & average contact days         & 0.500    & 0.706                   \\ \hline
Gini coefficient of CP score & average contact days         & 0.339    & 0.582                   \\ \hline
average degree               & inpatient days in hospital   & 0.565    & 0.751                   \\ \hline
Gini-outdeg                  & expense-per-inpatient-day    & 0.429    & 0.655                   \\ \hline
global cluster coefficient   & \#physicians                 & 0.358    & -0.598                  \\ \hline
\end{tabular}
\end{table}

\begin{table}[]
\centering
\vspace{0.3in}
\caption{Stability of estimated effects of network predictors for four network features: average-degree (f1), (out-, out-)-assortativity in induced-network (f25) and Gini coefficient of component size distribution in induced network (f16). The changes in one predictor when others are added reflects the collinearity between the predictors.}
\vspace{0.02in}
\label{table:coefficient-fixed-effects}
\begin{tabular}{|l|l|l|}
\hline
Model & Selected Network Features        & Coefficient ($\beta_{1}$)                                          \\ \hline
m1    & f1                            & 19.11                                                 \\ \hline
m2    & (f1, f25)                     & (16.03, 8,56)                                   \\ \hline
m3    & (f1, f25, f16)                & (14.70, 6.92, 4.08)                             \\ \hline
\end{tabular}
\end{table}

\begin{table}[]
\scriptsize
\centering
\vspace{-0.05in}
\caption{Associations between healthcare attributes and network features. f1 (average degree). A full list of potential predictors/network features is in Table~\ref{Appendix-predictors}. The features with an asterisk also have significant interactions with time for the corresponding healthcare attribute. The coefficients of interaction items are in Table~\ref{Appendix-interaction-coefficients}.}
\vspace{0.05in}
\label{tab:MLR-more-results}
\begin{tabular}{|l|l|l|}
\hline
Healthcare Attribute               & Selected Network Features          & Corresponding Coefficients              \\ \hline
medical-discharge-per1000-people        & (f1, f25, f16)                & (14.70, 6.92, 4.08) \\ \hline
hospital-admissions-per1000-decedents  & (f1, f22, f25, f$16^{*}$)            & (84.04, 16.58, 15.22, 17.99)        \\ \hline
long-term-opioids-receipt          & (f1, f7, f20, f26, f24)            & (0.70, -0.077, -0.32, 1.30, 0.82)       \\ \hline
general-inpatient-days-in-hospital & (f$1^{*}$, f25, f18)                     & (0.14, 0.050, 0.0051)    \\ \hline
medical reimbursement per capita      & (f23, f$29^{*}$, f30)                 & (-1.6e-5, 4.1e-5, -1.4e-3) \\ \hline
Gross State Product (GSP)          & (f$9^{*}$, f22, f$17^{*}$, f1, f3)         & (459892.4, -79067.3, -21304.8, -8886.8, -10331.9)  \\ \hline
\#medicare-beneficiaries            & (f$9^{*}$, f5)                           & (974692.9, -59712.4)    \\ \hline
\#nodes-per-1000-people             & (f$1^{*}$, f6, f19)                      & (0.16, 0.023, -0.15) \\ \hline
mortality                          & (f1, f16, f10)                     & (0.12, -0.038, -0.12)      \\ \hline
expense-per-inpatient-day          & (f$5^{*}$, f7, f9)                       & (125.63, -85.44, 55.05)     \\ \hline
\#hospital-beds-per-1000-people     & (f7, f$1^{*}$, f$5^{*}$, f16, f$26^{*}$, f29)        & (0.048, 0.44, -0.13, -0.11, 0.39, -0.012) \\ \hline

\end{tabular} 
\end{table}
\begin{table}[]
\centering
\vspace{-0.07in}
\caption{Full list of network features for the mixed-effect (hierarchical or multi-level) regression model.}
\vspace{0.05in}
\label{Appendix-predictors}
\begin{tabular}{|l|l|}
\hline
Feature ID & Name  \\ \hline
f1         & average degree of intra-state network  \\ \hline
f2         & $\alpha$ of in-degree Power Law of intra-state network  \\ \hline 
f3         & Gini coefficient of in-degree distribution of intra-state network \\ \hline
f4         &  $\alpha$ of out-degree Power Law of intra-state network    \\ \hline
f5         &  Gini coefficient of out-degree distribution of intra-state network  \\ \hline
f6         &  diameter of intra-state network    \\ \hline
f7         &  global clustering coefficient of intra-state network    \\ \hline
f8         &  local clustering coefficient of intra-state network   \\ \hline
f9         &  \#nodes in intra-state network     \\ \hline
f10         & \#edges in intra-state network     \\ \hline
f11         &  undirected assortativity of intra-state network    \\ \hline
f12         &   (in-,in-)assortativity of intra-state network    \\ \hline
f13         &  (out-,out-) assortativity of intra-state network    \\ \hline
f14         &   (in-,out-)assortativity of intra-state network   \\ \hline
f15         &   (out-,in-)assortativity of intra-state network   \\ \hline
f16         &  Gini coefficient of component size distribution of induced network    \\ \hline
f17         &  size of dominant component of induced network    \\ \hline
f18         &  diameter of induced network    \\ \hline
f19         &  global clustering coefficient of induced network     \\ \hline
f20         &  local clustering coefficient of induced network    \\ \hline
f21         &    \#nodes in induced network    \\ \hline
f22         &  \#edges of induced network    \\ \hline
f23         &  undirected assortativity of induced network    \\ \hline
f24         & (in-,in-)assortativity of induced network     \\ \hline
f25         &  (out-,out-) assortativity of induced network    \\ \hline
f26         &  (in-,out-)assortativity of induced network    \\ \hline
f27         &  (out-,in-)assortativity of induced network    \\ \hline
f28         &  Gini coefficient of CP scores for all nodes in a state    \\ \hline
f29         &  entropy of the core node distribution under different parameter settings    \\ \hline
f30         &  \#states that the core node in terms of CP score can directely reach out    \\ \hline
f31         &  \#cross-state referrals of the core node in terms 1.0 CP score in 2009-2014    \\ \hline

\end{tabular}

\end{table}

\begin{table}[]
\scriptsize
\centering
\vspace{0.3in}
\caption{Coefficients of significant interaction terms with linear time (in units of years) in the mixed-effect regression model.}
\label{Appendix-interaction-coefficients}
\begin{tabular}{|l|l|l|l|}
\hline
Healthcare Attribute &                 Network Feature  & Time (year) & Coefficients of Interaction Items ($\beta_{2}$)\\ \hline
hospital-admissions-per1000-decedents     &    f16                  & 2010-2012         &   (5.35, 9.54, 0.52)              \\ \hline
general-inpatient-days-in-hospital        &    f1                   & 2010-2012         &   (-0.0072,-0.019,-0.040)             \\ \hline
medical reimbursement per capita          &    f29                  & 2010-2012       &  (-2.2e-05, 8.7e-06, -4.9-05)              \\ \hline
Gross State Product (GSP)                 &    f9                   & 2010-2012          &  (29883.2, 55874.3, 88033.6)             \\ \hline
Gross State Product (GSP)                 &    f17                  & 2010-2012          &  (-24454.9, -44870.2, -64875.2)            \\ \hline
\#medicare-beneficiaries                  &    f9                   & 2010-2012         &  (10274.3, 20157.3, 35119.1)               \\ \hline
\#nodes-per-1000-people                   &     f1                  & 2010-2014         &  (0.030, 0.042,0.065,0.090,0.098)                  \\ \hline
expense-per-inpatient-day                 &     f5                  & 2010-2014         &  (4.0, 22.5, 41.1, 46.7, 64.5)                 \\ \hline
\#hospital-beds-per-1000-people           &     f1                  & 2010-2014         &  (-0.22, -0.19, -0.19, -0.15, -0.17)           \\ \hline
\#hospital-beds-per-1000-people           &     f5                  & 2010-2014         &  (-0.12, -0.11, -0.10, -0.10, -0.10)              \\ \hline
\#hospital-beds-per-1000-people           &     f26                 & 2010-2014         &  (-0.30, -0.27, -0.30, -0.30, -0.32)               \\ \hline
\end{tabular}
\end{table}
\end{document}